\documentclass[a4paper,USenglish,cleveref,autoref,thm-restate]{lipics-v2021}

\hideLIPIcs
\title{Mimicking Networks for Constrained Multicuts in Hypergraphs} 

\author{Kyungjin Cho}{Department of Computer Science and Engineering, POSTECH, Pohang, Korea}{kyungjincho@postech.ac.kr}{0000-0003-2223-4273}{} 
 \author{Eunjin Oh}{Department of Computer Science and Engineering, POSTECH, Pohang, Korea}{eunjin.oh@postech.ac.kr}{0000-0003-0798-2580}{} 

\authorrunning{K. Cho and E. Oh} 

\Copyright{Kyungjin Cho and Eunjin Oh} 



\category{} 

\relatedversion{} 

\funding{This work was partly supported by Institute of Information \& Communications Technology Planning \& Evaluation (IITP) grant funded by the Korea government (MSIT) (No.RS-2024-00440239, Sublinear Scalable Algorithms for Large-Scale Data Analysis) and the National Research Foundation of Korea (NRF) grant funded by the Korea government (MSIT) (No.RS-2024-00358505 and No.RS-2024-00410835).}

\nolinenumbers 

\usepackage{thm-restate}
\usepackage{algorithm}
\usepackage{algpseudocode}
\usepackage{todonotes}
\EventEditors{John Q. Open and Joan R. Access}
\EventNoEds{2}
\EventLongTitle{42nd Conference on Very Important Topics (CVIT 2016)}
\EventShortTitle{CVIT 2016}
\EventAcronym{CVIT}
\EventYear{2016}
\EventDate{December 24--27, 2016}
\EventLocation{Little Whinging, United Kingdom}
\EventLogo{}
\SeriesVolume{42}
\ArticleNo{23}

 \newcommand{\mincut}{\textsf{min-cut}}
 \newcommand{\capacity}{\textsf{cap}}

   \newcommand{\sink}{\textsf{sink}}
   \newcommand{\smal}{\textsf{small}}

\begin{document}
\maketitle
\begin{abstract}    
    In this paper, we study a \emph{multicut-mimicking network} for a hypergraph over terminals $T$ with a parameter $c$.
    It is a hypergraph preserving the minimum multicut values of any set of pairs over $T$ where the value is at most $c$.
    This is a new variant of the multicut-mimicking network of a graph in [Wahlstr{\"o}m ICALP'20], which introduces a parameter $c$ and extends it to handle hypergraphs.
    Additionally, it is a natural extension of the \emph{connectivity-$c$ mimicking network} introduced by [Chalermsook et al. SODA'21] and [Jiang et al. ESA'22] that is a (hyper)graph preserving the minimum cut values between two subsets of terminals where the value is at most $c$.

    We propose an algorithm for a hypergraph that returns a multicut-mimicking network over terminals $T$ with a parameter $c$ having $|T|c^{O(r\log c)}$ hyperedges in $p^{1+o(1)}+|T|(c^r\log n)^{\tilde{O}(rc)}m$ time, where $p$ and $r$ are the total size and the rank, respectively, of the hypergraph. 
\end{abstract}

\section{Introduction}
{Graph sparsification} is a fundamental tool in theoretical computer science. By reducing the size of a graph while preserving specific properties, such as the value of an objective function or its approximation, graph sparsification significantly enhances computational efficiency. This is particularly crucial for practical applications with limited resources and for handling large-scale data in real-world problems. Due to these advantages, various types of sparsification results have been presented over the decades, including spanners~\cite{chechik2018near,filtser2016greedy}, flow sparsification~\cite{chen20241+,krauthgamer2023exact}, and cut sparsification~\cite{chekuri2018minimum}. Additionally, their applications have been widely studied, such as in designing dynamic algorithms~\cite{jin2022fully}.
In this paper, we focus on graph sparsification specifically tailored for hypergraph separation and cut problems. 

Hypergraph separation and cut problems have garnered significant attention due to their extensive applications and theoretical challenges. These problems are particularly compelling because hypergraphs offer more accurate modeling of many complex real-world scenarios compared to normal graphs. Examples include VLSI layout~\cite{alpert1995recent}, data-pattern-based clustering~\cite{ozdal2004hypergraph}, and social tagging networks~\cite{zhang2010hypergraph}. The transition from graph to hypergraph separation problems opens up new avenues for research, driven by the need to address the unique properties and complexities inherent in hypergraphs. Researchers have thus increasingly focused on developing graph algorithms and theoretical frameworks for hypergraph problems, such as the small set expansion problem in hypergraphs~\cite{louis2014approximation}, spectral sparsification in hypergraphs~\cite{bansal2019new,kapralov2022spectral}, and connectivity-$c$ mimicking problem in hypergraphs~\cite{jiang2022hypergraphvertex}. This growing interest underscores the critical importance of hypergraph separation and cut problems in both theoretical and practical applications.

One of the key problems in (hyper)graph sparsification is the \emph{mimicking problem}.
It aims to find a graph that preserves minimum cut sizes between any two subsets of vertices called \emph{terminals}.
A \emph{cut} between two sets of vertices is a set of edges whose removal disconnects the given two sets.
Kratsch et al.~\cite{kratsch2012representative} showed that there is a mimicking network with $O(\tau^{3})$ edges, where $\tau$ is the number of edges incident to terminals.
Chalermsook et al.~\cite{soda2021vertexspar} introduced a \emph{constraint version}, called \emph{connectivity-$c$ mimicking problem}, that aims to preserve minimum cut sizes between every two subsets of terminals where the size is at most $c$, and they showed that there is such a graph with $O(kc^4)$ edges, which was later improved to $O(kc^3)$~\cite{liu2023vertex_poly}, where $k$ is the number of terminals.
This result was extended to hypergraphs by Jiang et al.~\cite{jiang2022hypergraphvertex}.

A crucial variant of the mimicking problem is the \emph{multicut-mimicking problem}.
A \emph{multicut} of pairs of vertices is a set of (hyper)edges whose removal disconnects each given pair.
Studies have shown that the multicut problem is highly beneficial in various applications, including network design, optimization, and security, where maintaining specific connectivity while minimizing resources is necessary compared to cut problems~\cite{kappes2016higher}. 
It is already known that the problem is NP-hard even for graphs~\cite{dahlhaus1994complexity,marx2006parameterized}.
A \emph{multicut-mimicking network} for a set of terminals in a (hyper)graph is a (hyper)graph that preserves the size of the minimum multicut for any set of pairs of terminals.  
Kratsch et al.~\cite{kratsch2012representative} proposed a method to obtain a multicut-mimicking network by contracting edges in 
a graph except at most $\tau^{O(k)}$ edges, where 
$k$ and $\tau$ are the numbers of terminals and incident edges to terminals, respectively. 
Wahlstr{\"o}m~\cite{wahlstrom2022quasipolynomial_multicut} refined this method and reduced the number of edges to $\tau^{O(\log \tau)}$.

Unlike the mimicking problem, there is no existing result for the constraint version of \emph{multicut-mimicking network problem}, even for graphs. 
We further study the multicut-mimicking problem by introducing a parameter $c$.
Precisely, we present an algorithm to compute a hypergraph, that preserves the size of the minimum multicut for any set of pairs of terminals where the size is at most $c$, with a linear size in the number of terminals while the previous best-known result for multicut-mimicking network, without the parameter $c$, has an exponential size~\cite{kratsch2012representative}.
It will allow for more refined control over the sparsification process, enabling the construction of smaller and more efficient networks.
For instance, this notion in {mimicking problem} was utilized for a dynamic connectivity problem~\cite {jin2022fully}.

\subparagraph{Our result.}
In this paper, we study \emph{vertex sparsifiers for multiway connectivity} with a parameter $c > 0$. 
Our instance $(G, T,c)$ consists of an undirected hypergraph $G$, terminal set $T \subseteq V(G)$, and a parameter $c$. 
Precisely, we construct a hypergraph that preserves minimum multicut values over $T$ where the value is at most $c$.
It is the first result for the multicut-mimicking networks adapting the parameter $c$ even for graphs.

Previously, the best-known multicut-mimicking network had a quasipolynomial size in the total degree of terminals in $T$~\cite{wahlstrom2022quasipolynomial_multicut}, specifically $|\partial T|^{O(\log |\partial T|)}$. 
By introducing the parameter $c$, we demonstrate that a multicut-mimicking network for 
$(G,T,c)$ exists with a size linear in $|T|$. 
This allows us to utilize the near-linear time framework of Jiang et al.~\cite{jiang2022hypergraphvertex} to find a mimicking network using the \emph{expander decomposition} of Long and Saranurak~\cite{long2022near}.
Our result is summarized in Theorem~\ref{thm:near_linear_algo}. 
Here, $m=|E(G)|$ and $r$ is the rank of $G$. 

\begin{restatable}{theorem}{thmNearLinear}\label{thm:near_linear_algo}
For $(G, T,c)$, we can compute a multicut-mimicking network of at most $kc^{O(r\log c)}$ hyperedges in {$p^{1+o(1)}+k(c^{r\log c}\log n)^{O(rc)} m$ time}, where $k=|T|$ and $p=\sum_{e\in E}|e|$.
\end{restatable}

\subparagraph{Outline.}
Our work extends the framework from connectivity-$c$ mimicking networks for hypergraphs, introduced by Jiang et al.~\cite{jiang2022hypergraphvertex}, to multicut-mimicking networks, as well as adapting methods from multicut-mimicking networks for graphs, introduced by Wahlstr{\"o}m~\cite{wahlstrom2022quasipolynomial_multicut}, to hypergraphs with a parameter 
$c$. While we broadly follow the previous approaches, we extend the concepts and methods used in the previous studies to fit the multicut-mimicking problem in hypergraphs with the parameter 
$c$. This extension allows us to handle the complexities of hypergraphs effectively.

We introduce notions used in this paper in Section~\ref{sec:prel} and illustrate an efficient algorithm to compute a small-sized multicut-mimicking network outlined in Theorem~\ref{thm:near_linear_algo} in Section~\ref{sec:near-linear}. 
We give an upper bound for the size of minimal multicut-mimicking networks of hypergraphs in Section~\ref{sec:exists}, which is a witness for the performances of our algorithm outlined in Theorem~\ref{thm:near_linear_algo}. 

\section{Preliminaries}\label{sec:prel}
A \emph{hypergraph} $G$ is a pair $(V(G), E(G))$, where $V(G)$ denotes the set of vertices and $E(G)$ is a collection of subsets of $V(G)$ referred to as \emph{hyperedges}.
If the context is clear, we write $V$ and $E$. 
The \emph{rank} of $G$ is defined as the size of its largest hyperedge. 
For a vertex $v\in V$, a hyperedge $e$ is said to be \emph{incident} to $v$ if $v\in e$.
For a vertex set $X \subset V$, let $\partial_G X$ denote the set of hyperedges in $E$ containing at least one vertex from $X$ and one from $V \setminus X$, and let $E(X)$ denote the set of hyperedges fully contained in $X$. 
Additionally, we let $G/e$ denote the \emph{contraction} of a hyperedge $e$ in $G$ obtained by merging all vertices in $e$ into a single vertex and modifying the other hyperedges accordingly. 
A \emph{path} in a hypergraph is defined as a sequence of hyperedges such that any two consecutive hyperedges contain a common vertex.

Consider a \emph{partition} $(X_1, \dots, X_s)$ of a vertex set $X \subseteq V$.
We call each subset $X_i$ a \emph{component} of this partition.
Additionally, the \emph{cut} of $(X_1, \dots, X_s)$ in $G$ is defined as the set of hyperedges of $G$ intersecting two different components.
We let $[a]=\{1,\dots,a\}$ and $[a,b]=\{a,\dots, b\}$ for integers $a<b$. 
Furthermore, let $|X|$ denote the number of elements of a set $X$.
For a hyperedge $e\in E(G)$, we let $|e|$ to denote the number of vertices in $e$.

In this paper, an instance $(G, T, c)$ consists of a hypergraph $G$, 
a set $T \subseteq V(G)$, and a positive constant $c$. We refer to the vertices in $T$ as \emph{terminals}.
For a set $R$ of pairs of $T$, a \emph{multicut} of $R$ in $G$ is a set of hyperedges $F\subset E(G)$ such that every connected component in $G\setminus F$ contains at most one element of every pair $\{t,t'\}\in R$.
We construct a \emph{multicut-mimicking network} $H$ of $(G,T,c)$ that is a hypergraph obtained from $G$ by contraction of hyperedges which preserves the size of minimum {multicut} for all set $R$ of pairs over $T$ where the size is at most $c$.
Precisely, if a multicut $F$ of $R$ exists in $G$ with $|F| \leq c$, then a multicut $F'$ of $R$ exists in $H$ with $|F'|\leq |F|$.  
We say $H$ is \emph{minimal} if no contraction $H/e$ is a multicut-mimicking network of $(G, T,c)$.
Analogously, we define a \emph{minimal instance}.
We address the multicut-mimicking problem by utilizing \emph{multiway cuts}.

\subsection{Multiway Cuts and Essential Edges}\label{sec:multiwaycut} 

We refer to a partition of terminals $T$ as a \emph{terminal partition}. 
For a terminal partition $\mathcal T$, a hyperedge set $F$ is termed \emph{a multiway cut of $\mathcal T$} if any two terminals from different components in $\mathcal T$ are not connected in $G\setminus F$. 
Furthermore, if there is no multiway cut of size less than $|F|$, then $F$ is called a \emph{minimum multiway cut of $\mathcal T$ in $G$}. Let $\mincut_G(\mathcal T)$ denote the minimum multiway cut size of the partition in $G$.
A hyperedge $e\in E(G)$ is said to be \emph{essential} for $(G,T,c)$ if there exists a terminal partition $\mathcal{T}$ with $\mincut_G(\mathcal{T})\leq c$ such that every minimum multiway cut of $\mathcal{T}$ in $G$ contains $e$. Otherwise, $e$ is \emph{non-essential}. 

Multicuts and multiway cuts in graphs are closely related~\cite[Proposition 2.2]{wahlstrom2022quasipolynomial_multicut}. We observe that this close relation also holds in hypergraphs.
Briefly, a contraction of a non-essential hyperedge is a multicut-mimicking network by Lemma~\ref{lem:contracting}.


\begin{restatable}{lemma}{Contracting}\label{lem:contracting}
    For a hyperedge $e$ of $G$, $G/e$ is a multicut-mimicking network for $(G, T,c)$ if and only if $e$ is non-essential for $(G, T,c)$.
\end{restatable}
\begin{proof}       
    If $e$ is essential for $(G,T,c)$, then $\mincut_{G/e}(\mathcal T)>\mincut_G(\mathcal T)$ for some terminal partition $\mathcal T$ where any minimum multiway cut of $\mathcal T$ for $T$ contains $e$ and $\mincut_G(\mathcal T)\leq c$.
    Let $R$ be the set of all pairs of $T$ where two elements are from different components in $\mathcal T$.
    Any multicut of $R$ is a multiway cut of $\mathcal T$, and vice versa.
    Thus the minimum multicut value of $R$ in $G/e$ is strictly larger than the value in $G$.
    Thus, the `if' direction holds. 

    To prove the `only if' direction, 
    we assume that $e$ is non-essential, then fix an arbitrary set $R$ of pairs of $T$ which has a minimum multicut $F$ in $G$ of size at most $c$.
   Let $\mathcal T$ be the terminal partition according to $G\setminus F$.
    Any multiway cut $\mathcal T$ is a multicut of $R$ and $\mincut_G(\mathcal T)\leq |F|$.
    There is a multiway cut of $\mathcal T$ in $G$ excluding $e$ of size at most $|F|$ since $e$ is non-essential. 
    Therefore, there is a minimum multiway cut of $\mathcal T$ in $G/e$ of size at most $|F|$ which is also a minimum multicut of $R$ in $G/e$.
    This implies that the minimum multicut value of $R$ in $G/e$ cannot exceed that in $G$.
    Since contracting a hyperedge cannot decrease any minimum multicut value, $G/e$ preserves the minimum multicut value of $R$.
    In conclusion, $G/e$ is a multicut-mimicking network.
\end{proof}

\begin{figure}[t]
	\begin{center}
		\includegraphics[width=0.6\textwidth]{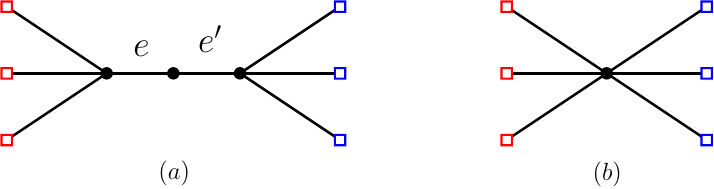}
		\caption{\label{fig:no_multiple_cont}\small
		  The colored square points mark terminals in the graph. (a) The edges $e$ and $e'$ are both non-essential edges.
            (b) Contraction $\{e,e'\}$ cannot preserve the minimum cut between red terminals and blue terminals.
                            \vspace{-1em}
		}
	\end{center}
\end{figure}

A multicut-mimicking network is minimal if and only if every hyperedge is essential.
Note that we cannot contract multiple non-essential hyperedges simultaneously. 
This is because even if two hyperedges $e$ and $e'$ are non-essential in $(G, T,c)$, the contraction $G/\{e,e'\}$ might not be a multicut-mimicking network of $(G, T,c)$ even for a graph $G$, not a hypergraph. Figure~\ref{fig:no_multiple_cont} shows a counterexample. 
In this paper, we construct a multicut-mimicking network by finding and contracting non-essential hyperedges one by one.

\subsection{Restricted Hypergraphs and Subinstances}\label{sec:restrictedhyper}
\begin{figure}
	\begin{center}
		\includegraphics[width=0.85\textwidth]{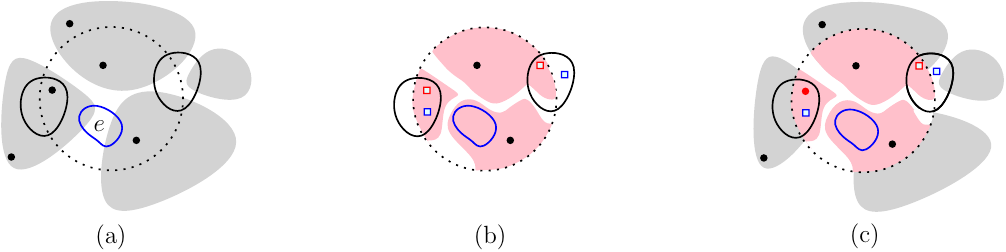}
		\caption{\label{fig:subinstance}\small
		  Illustration of proof of Lemma~\ref{lem:essential_in_subinstance}.
		  (a) Illustration of a terminal partition $\mathcal T$ and $G\setminus F$. The dotted circle separates $X$ (inside) and $V\setminus X$ (outside).
            (b) Illustration of $\hat G[X]$, terminal partition $\mathcal T_X$, and the vertex partition consisting of $\hat G[X]\setminus F_X$.
            Blue and red squared points are anchored terminals $a_{e'}$ and $t_{e'}$, respectively, for $e'\in \partial X$.
            (c) Illustration of $G\setminus (F_X\cup \bar F)$. The multiway cut $(F_X\cup \bar F)$ is a minimum multiway cut of $\mathcal T$ and excluding $e$.
                                        \vspace{-1em}
		}
	\end{center}
\end{figure}

The \emph{subinstance} $(\hat G[X], T_{X}, c_X)$ of $(G,T,c)$ for $X\subset V(G)$ is constructed as follows. Refer to Figure~\ref{fig:subinstance} (a-b).
For each hyperedge $e\in \partial X$, we insert a vertex $a_e$, and we choose an arbitrary terminal $t_e$ in $e\cap (X\cap T)$. 
If no such terminal exists, we insert a new vertex $t_e$. 
We refer to $a_e, t_e$ as the \emph{anchored terminals} of $e$, and $(e\cap X)\cup\{a_e,t_e\}$ as the \emph{restricted hyperedge} of $e$, denoted by $e_X$.
We obtain $\hat G[X]$ from $G$ through the following:
\textsf{i)} add the anchored terminals of the hyperedges in $\partial X$,
\textsf{ii)} replace the hyperedges in $\partial X$ with their restricted hyperedges, and
\textsf{iii)} delete the vertices $V\setminus X$ and the hyperedges $E(V\setminus X)$. We call it the \emph{restricted hypergraph} of $G$ for $X$.
Let $T_X$ denote the set of all terminals in $T \cap X$ and the anchored terminals, and  $c_X=\min \{c,|T_X|\}$. 
All subinstances preserve all essential hyperedges in the original instance by Lemma~\ref{lem:essential_in_subinstance}.
Figure~\ref{fig:subinstance} sketches its proof.

\begin{restatable}{lemma}{LemEssentialSubinstance}\label{lem:essential_in_subinstance}
    If a hyperedge $e$ is non-essential in a subinstance $(\hat G[X],T_X,c_X)$, then $e$ is also non-essential in the original instance $(G,T,c)$. 
    Furthermore, $e$ is not in $\partial_G X$.
\end{restatable}
\begin{proof}
    Note that $e$ is not a restricted hyperedge since a restricted hyperedge $e'_X$ is the unique cut separating two anchored terminals $a_{e'}$ and $t_{e'}$ within $e'_X$.    
    Thus, $e$ is in $E(X)\subset E(G)$. We fix an arbitrary terminal partition $\mathcal T$ of $T$ with $\mincut_G(\mathcal T)\leq c$, and we construct a minimum multiway cut of $\mathcal T$ excluding $e$. 
    Let $F$ be a minimum multiway cut of $\mathcal T$ in $G$, and let $\bar F=F\setminus (E(X)\cup \partial X)$.
    Note that $F\setminus \bar F$ is the set of hyperedges in the multiway cut $F$ intersecting $X$.
    Note that if $F$ has the hyperedge $e$, then $e$ is in $F\setminus \bar F$.

    We obtain a multiway cut of $\mathcal T$ from $X$ by replacing $F\setminus \bar F$ as a multiway cut excluding $e$ using the subinstance $(\hat G[X],T_X,c_X)$.
    Recall that $T_X$ is the union of $T\cap X$ and the anchored terminals $a_{e'}$ and $t_{e'}$ for each restricted hyperedges $e'_X$ with $e'\in \partial X$.
    Furthermore, $a_e$'s are not in the original graph $G$.
    Let $\mathcal T_X$ be the terminal partition of $T_X$ constructed as follows.
    First, we start from the partition $\mathcal T_X$ of $T\cap X$ according to $G\setminus F$.
    Then we insert $t_{e'}$ for each $e'_X\in \partial X$ into an arbitrary component in $\mathcal T_X$ which is connected in $G\setminus F$ if $t_{e'}\notin T\cap X$.
    Finally, we insert $a_{e'}$ for each $e'_X\in \partial X$ into the same component in $\mathcal T_X$ with $t_{e'}$ if $e'\notin F$, otherwise, we insert the isolated component $\{a_{e'}\}$ into $\mathcal T_X$.
    By construction, the restricted hyperedges of $F$ on $X$ form a multiway cut of the terminal partition $\mathcal T_X$ of $T_X$ of which the size is at most $c_X=\min \{c, |T_X|\}$.
    Since $e$ is not an essential hyperedge $(\hat G[X], T_X, c_X)$, there is a minimum multiway cut of $\mathcal T_X$ in $\hat G[X]$ excluding $e$.
    Let $F_X$ be the set of hyperedges $e'$ in the original graph $G$ where its restricted hyperedge $e'_X$ is in the minimum multiway cut of $\mathcal T_X$ excluding $e$.

    We demonstrate that $F_X\cup \bar F$ is a multiway cut of $\mathcal T$ in $G$ excluding $e$.
    This completes the proof since it excludes $e$ and its size is at most that of $F$ which is a minimum multiway cut of $\mathcal T$.
    For this, we fix an arbitrary path $\pi$ in $G$ between two components of $\mathcal T$, and we show that it contains at least one hyperedge in $F_X\cup \bar F$.
    Recall that $F$ is a multiway cut of $\mathcal T$, and thus, $\pi$ has a hyperedge in $F$.
    If $\pi$ is fully contained in $\hat G[X]$ or $\hat G[V\setminus X]$, our claim holds naturally.
    Furthermore, if $\pi\ni e'$ with $e'\in F \cap \partial X$, then the anchored terminals $a_{e'}$ and $t_{e'}$ are in the different components in $\mathcal T_X$ by construction. 
    Thus, $e'$ is a hyperedge in $F_X$, which implies our claim holds.
    In the other scenario, $\pi$ is decomposed into several subpaths by the hyperedges $\partial X$.
    If a subpath fully contained in $\hat G[V\setminus X]$ has a hyperedge in $F$, the hyperedge is in $\bar F$.
    For a subpath fully contained in $\hat G[X]$, the subpath is not in $G\setminus F_X$ if it is not in $G\setminus F$ since $F_X$ corresponds to the multiway cut of $\mathcal T_X$ of $T_X$ according to $G\setminus F$.
    In conclusion, $\pi$ has a hyperedge in $F_X\cup \bar F$.
    Thus, $F_X\cup \bar F$ is a multiway cut of $\mathcal T$ in $G$. 
\end{proof}

\section{Efficient Algorithm for Computing Multicut-Mimicking Networks}\label{sec:near-linear}
In this section, we design an algorithm to compute a minimal multicut-mimicking network for $(G,T,c)$, where $G$ is a hypergraph. 
We broadly follow the approach of Jiang et al.~\cite{jiang2022hypergraphvertex}. Since their original algorithm was designed for mimicking networks, not multicut-mimicking networks, we need to modify their algorithm. First, we introduce their algorithm briefly.

Jiang et al.~\cite{jiang2022hypergraphvertex} designed an algorithm to find a connectivity-$c$ mimicking network for hypergraphs using the \emph{expander decomposition} of Long and Saranurak~\cite{long2022near}.
Precisely, they designed an algorithm to find a connectivity-$c$ mimicking network of size linear in $|T|$ for an \emph{expander} $G$ with terminals $T$, and then, they extended it for a general hypergraph using the expander decomposition.
For a parameter $\phi>0$, a hypergraph $G$ (and instance $(G,T,c)$)  is called a \emph{$\phi$-expander} if either $E(X)$ or $E(V\setminus X)$ has at most $\phi^{-1}|\partial X|$ hyperedges for any vertex set $X\subset V(G)$.  The following explains the key idea of their and our algorithm.

Recall that contracting a non-essential hyperedge obtains a smaller mimicking network by Lemma~\ref{lem:contracting}.
Generally, a non-essential hyperedge can be found by comparing every terminal partition and subset of hyperedges, which is time-consuming. However, if we suppose that the given instance is an \emph{expander}, then we can do this more efficiently by comparing \emph{useful terminal partitions} and their minimum multiway cuts instead of whole partitions and hyperedge subsets.
We adapt concepts used in the previous research such as \emph{useful terminal partitions} to suit our needs, and we newly introduce the concept \emph{core} of a multiway cut which refers to a small-sized vertex set including whole hyperedges of the multiway cut.

\subparagraph*{Useful terminal partitions, connected multiway cuts, and cores.}
Assume that the instance $(G, T,c)$ is a $\phi$-expander and $G$ is a connected hypergraph. 
For a multiway cut $F$ in $G$, we define the \emph{core} of $F$ as the union $C$ of connected components $X$ in $G\setminus F$ with $|E(X)|\leq \phi^{-1}|F|$. 
The definition of an expander guarantees that at most one component in $G\setminus F$ has more than $\phi^{-1}|F|$ hyperedges. 
Therefore, the multiway cut $F$ includes all hyperedges $\partial C$ and is contained in $E(C)\cup \partial C$.
We say $F$ is a \emph{connected multiway cut} in $(G, T,c)$ if it is a minimum multiway cut of some terminal partition with $|F|\leq c$ and $T\cap C$ is connected in $\hat G[C]$, where $C$ is the core of $F$ and $\hat G[C]$ is the restricted hypergraph defined in Section~\ref{sec:restrictedhyper}.
A terminal partition is said \emph{useful} if every minimum multiway cut of it is a connected multiway cut.

Since every core of connected multiway cuts has a small number of vertices and hyperedges in $\phi$-expander, we can enumerate all of them efficiently. 
Additionally, since a core includes its corresponding multiway cut, we can also enumerate all connected multiway cuts and useful partitions. Details are in Section~\ref{sec:expander}.
The most interesting property is that comparing all useful partitions is sufficient to find a non-essential hyperedge.

\begin{figure}
	\begin{center}
		\includegraphics[width=0.9\textwidth]{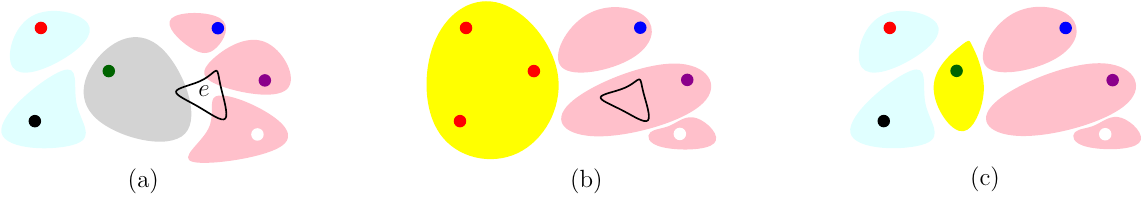}
		\caption{\label{fig:useful_essential}\small
		  Illustration of the proof of Lemma~\ref{lem:useful_essential}. 
    (a) Illustration of the terminal partition $\mathcal T$ and the vertex partition according to $G\setminus F$.
    The middle gray area is $X$.
    The right three red areas form $C$.
    (b) Illustration of the terminals partition $\mathcal T'$ and the partition of $G\setminus F'$.
    (c) Illustration of the partition $G\setminus (F'\cup \bar F)$.
    $F'\cup \bar F$ is a multiway cut of $\mathcal T$ excluding $e$.
                            \vspace{-2.5em}
		}
	\end{center}
\end{figure}

\begin{lemma}\label{lem:useful_essential}
    A hyperedge $e\in E$ is essential for $(G, T,c)$ if and only if there is a useful partition such that every minimum multiway cut for it contains $e$.
\end{lemma}
\begin{proof}
    The `if' direction is trivial since it is consistent with the definition of essential. 
    For the `only if' direction, we assume that $e$ is essential for $(G, T, c)$.
    Let $\mathcal{T}$ be a terminal partition minimizing $\mincut_G(\mathcal{T})$ 
    of which every minimum multiway cut involves $e$. 
    In the following, we show that $\mathcal{T}$ is a \emph{useful partition} by contradiction. Figure~\ref{fig:useful_essential} illustrates this proof.

    Assume that $\mathcal{T}$ is not a useful partition for $(G, T, c)$, and let $F$ be a minimum multiway cut of $\mathcal{T}$ which is not a connected multiway cut. 
    If the core of $F$ is $V(G)$, then $F$ is a connected multiway cut.
    Therefore, the core is the complement of some connected component $X$ in $G\setminus F$.
    Let $C$ be the connected component in $G-X$ that intersects the hyperedge $e$. Refer to Figure~\ref{fig:useful_essential}(a).
    Then we decompose the multiway cut $F$ into $F_e$ and $\bar F$ where $F_e=F\cap E(C\cup X)$ and $\bar F=F\setminus F_e$.
    By the construction, we have $F_e\subsetneq F$ and $e\in F_e$.
    We construct a minimum multiway cut of $\mathcal T$ excluding $e$ that completes the proof.

    Let $\mathcal T'$ be the terminal partition according to $G\setminus F_e$.
    Since $F_e\subsetneq F$ and we chose $\mathcal T$ as minimizing $\mincut_G(\mathcal T)$ while any minimum multiway cut of $\mathcal T$ contains $e$, there is a multiway cut $F'$ of $\mathcal T'$ excluding $e$. 
    We claim that $F'\cup \bar F$ is a minimum multiway cut of $\mathcal T$ excluding $e$ which contradicts and completes the proof. Refer to Figure~\ref{fig:useful_essential}(b-c).
    Note that the multiway cut has a size at most $|F|$ and excludes $e$ by construction.
    Thus, it is sufficient to show that there is no path in $G\setminus (F'\cup \bar F)$ between two components in $\mathcal T$.

    Consider a path $\pi$ in $G$ between two terminals in different components in $\mathcal T$.
    Note that $\pi$ is not a path in $G\setminus F$, and thus, $\pi$ is not in $G\setminus F_e$ or $G\setminus \bar F$. 
    Recall that $F'$ is the multiway cut of the terminal partition according to $G\setminus F_e$.
    That means $\pi$ is not in $G\setminus F'$ if it is not in $G\setminus F_e$.
    Therefore, there is no path in $G\setminus (F'\cup \bar F)$ between two components in $\mathcal T$.
\end{proof}

\subsection{Useful Terminal Partitions in Expanders}\label{sec:expander}
In this section, we explain how to efficiently enumerate all useful terminal partitions and their minimum multiway cuts. Then, we explain how to compute a multicut-mimicking network for an expander using the enumerated list along with Lemma~\ref{lem:useful_essential}.
Here, the instance $(G, T,c)$ is a $\phi$-expander and $G$ is a connected hypergraph. 

The key is based on Observation~\ref{obs:property_connected_expander}. 
Briefly, cores in a $\phi$-expander have a small number of vertices and hyperedges. 
The observation enables us to find all cores of connected multiway cuts and subsequently enumerate all useful partitions. 
However, it is possible to enumerate terminal partitions that are not useful. 
Therefore, we need to \emph{prune} the enumerated lists for useful terminal partitions and their minimum multiway cuts.
\begin{restatable}{observation}{ObsConnectedExpander}\label{obs:property_connected_expander}
    For a connected multiway cut $F$ and its core $C$,
     the restricted hypergraph $\hat G[C]$ is connected and has at most $(3\phi^{-1}+1)|F|$ hyperedges.
\end{restatable}
\begin{proof}
    First, we show that $\hat G[C]$ has at most $(3\phi^{-1}+1)|F|$ hyperedges
    Since we assume that $G$ is a $\phi$-expander, at most one connected component in $G\setminus F$ has more than $\phi^{-1}|F|$ hyperedges within.
    If there is such a component $X$, then the core $C$ is $V\setminus X$. Furthermore $\hat G[V\setminus X]$ has at most $\phi^{-1}|F|+|F|$ hyperedges
    since we have $|E(V\setminus X)|\leq \phi^{-1}|F|$ from the assumption that $G$ is a $\phi$-expander, and thus, the lemma holds.
    In the following, we suppose that every connected component has at most $\phi^{-1}|F|$ hyperedges, and let $(V_1,\dots, V_s)$ be the vertex partition consisting of all connected components in $G\setminus F$.
    
    Note that $E(G)$ is decomposed by $E(V_1),\dots,E(V_s)$, and $F$. 
    For the smallest index $j^*$ in $[s]$ with $\sum_{i\in[j^*]}|E(V_i)|> \phi^{-1}|F|$,  the following inequalities hold by the $\phi$-expander condition:

        $$\sum_{i\in[j^*-1]}|E(V_i)|\leq \phi^{-1}|F|\quad \textnormal{ and }\sum_{i\in[j^*+1,s]}|E(V_i)|\leq \phi^{-1}|F|.$$
    Thus, $G$ has at most $(3\phi^{-1}+1)|F|$ hyperedges since the size of $E(V_{j^*})$ is at most $\phi^{-1}|F|$.

    We show that $\hat G[C]$ is connected.
    Recall that every terminal $T\cap C$ is connected in $\hat G[C]$.
    As observed before, $V\setminus C$ is connected in $G\setminus F$.
    This implies that the anchored terminals in $\hat G[C]$ are also connected to $T\cap C$.
    Precisely, we have $\partial C\subset F$ by construction.
    Additionally, for a hyperedge $e\in F$, if there is no path in $G\setminus (F\setminus e)$ between a vertex in $e$ and $T\cap C$, then we can obtain a smaller multiway cut $F'$ then $F$ so that the terminal partitions according to $G\setminus F$ and $G\setminus F'$, respectively, are equal.
    This contradicts that $F$ is a minimum multiway cut of some terminal partition, thus, this scenario does not happen.
    Furthermore, for a vertex in $\hat G[C]$ not connected to $T\cap C$, it is connected to an anchored terminal in $\hat G[C]$, and thus, it is connected to $T\cap C$.
    In conclusion, $\hat G[C]$ is a connected hypergraph.    
\end{proof}   

\subparagraph*{Enumerating connected multiway cuts.}
Jiang et al.~\cite{jiang2022hypergraphvertex} designed an algorithm to enumerate all connected vertex sets $C$ with $|\partial C|\leq c$, $|E(C)|\leq M$, and $t\in C$ in the $\phi$-expander $G$ when a vertex $t\in V(G)$ and two integers $c, M$ are given as an input. 
Algorithm~\ref{alg:2conncuts} is the pseudocode of the algorithm.
This algorithm takes $(r(M+c))^{O(rc)}$ time.
Additionally, the number of enumerated connected vertex sets is at most $(r(M+c))^{O(rc)}$.
We use this algorithm along with Observation~\ref{obs:property_connected_expander} for enumerating all connected multiway cuts. 

Note that every connected multiway cut has a core $C$ with $|E(C)|\leq (3\phi^{-1}+1) c$ by Observation~\ref{obs:property_connected_expander}. 
Furthermore, each of their core must contain a terminal $t\in T$ since a connected multiway cut is a minimum multiway cut of some terminal partition of $T$.
By utilizing the above algorithm for every terminal in $T$ by fixing two integers $c$ and $M=(3\phi^{-1}+1) c$, we can enumerate all cores $C$ for every connected multiway cut in  $(G, T,c)$.
Additionally, by enumerating all hyperedge sets $F'$ in $E(C)\cup (\partial_G C)$ with $|F'|\leq c$ and $\partial C\subset F'$, we can enumerate multiway cuts including all connected multiway cuts.
Since $E(C)$ has at most $(3\phi^{-1}+1) c$ hyperedges, each $C$ enumerates $(c\phi^{-1})^{O(c)}$ multiway cuts.
Algorithm~\ref{alg:multiway_conn} provides the pseudocode of this process. 

In conclusion, we enumerate $|T|(rc\phi^{-1})^{O(rc)}$ multiway cuts including all connected multiway cuts of size at most $c$ in $|T|(rc\phi^{-1})^{O(rc)}$ time.
Furthermore, there are at most $|T|(rc\phi^{-1})^{O(rc)}$ terminal partitions of $T$ contributed by the enumerated multiway cuts since at most $rc$ number of connected components are occurred by removing at most $c$ hyperedges.
These terminal partitions include all useful partitions since a minimum multiway cut of a useful partition is a connected multiway cut.
\begin{algorithm}[t]
\caption{ConnectedMultiwayCuts$(G,c,\phi)$}\label{alg:multiway_conn}
\hrule
\vspace{2pt}
\hspace*{\algorithmicindent} \textbf{Input:} A hypergraph $G$ who is $\phi$-expander and $c>0$\\
\hspace*{\algorithmicindent} \textbf{Output:} All connected multiway cuts $F\subset E(G)$ in $G$.
\vspace{2pt}
\hrule
\begin{algorithmic}[1]
\State $\mathcal C\gets \emptyset$
\For{each vertex $t\in V(G)$}
    \State $\mathcal C \gets \mathcal C \cup \textsf{EnumerateCutsHelp}(G,t,\emptyset\ ;c,3c\phi^{-1}+c)$  
\EndFor
\State $\mathcal F\gets \{\}$
\For {each $C\in \mathcal C$ and a hyperedge set $F'\subset E(C)$ with $|\partial C \cup F'|\leq c$}
    \State $\mathcal F \gets \mathcal F \cup \{\partial C \cup F'\}$
\EndFor
\State \textbf{return $\mathcal F$}
\end{algorithmic}
\end{algorithm}
\begin{algorithm}[t]
\caption{EnumerateCutsHelp$(G,t,X;c,M)$}\label{alg:2conncuts}
\hrule
\vspace{2pt}
\hspace*{\algorithmicindent} \textbf{Input:} A hypergraph $G$, a vertex $t$ in $G$, $X\subset V$, and two integers $c,M>0$\\
\hspace*{\algorithmicindent} \textbf{Output:} All connected vertex set $C$ in $G-X$ with $t\in C$, $|\partial_G C|\leq c$, $|E(C)|\leq M$.
\hrule
\begin{algorithmic}[1]

\If{$|X|> rc$}
    \State \textbf{return} $\{\}$
\Else
    \State {Run \textbf{DFS} in $\hat G[V\setminus X]$ starting from $t$ and \textbf{STOP} after visiting $M+1$ edges}
    \State $C \gets$ visited vertex set
    \State $\hat E \gets$ visited hyperedge set
 
    \If{$|\partial_G C|\leq c$ and $|\hat E|\leq M$} 
        \State \textbf{return} $C$
    \EndIf

    \State $\mathcal C\gets \{\}$
    \For{each $v\in C$}
        \State $\mathcal C\gets \mathcal C\cup \textsf{EnumerateCutsHelp}(G,t,X\cup \{v\};c,M)$
    \EndFor
    \State \textbf{return $\mathcal C$}
\EndIf
\end{algorithmic}
\end{algorithm}

\subparagraph*{Pruning useful terminal partitions.}\label{sec:deterministic_useful}
To check whether a terminal partition is useful or not, we need to verify if $C\cap T$ is connected in $\hat G[C]$ for each core $C$ of its minimum multiway cuts.
Specifically, it is sufficient to check the inclusion-wise minimal cores among them. 
For this, we utilize \emph{important cuts} $\partial R$, which inclusion-wise maximizes $R\subset V$ while maintaining the size of the cut $\partial R$.
The definition aligns with our needs as outlined in Lemma~\ref{lem:useful_checking}.

For two disjoint vertex sets $A$ and $B$, let $R$ be a vertex set containing $A$ while excluding $B$.
We say $\partial R$ is an \emph{important cut} of $(A,B)$ if there is no $R'\supsetneq R$ excluding $B$ with $|\partial R'|\leq |\partial R|$. 
This definition holds even if $A=\emptyset$.
In a directed graph, an important cut is defined analogously by setting $\partial R$ as the outgoing arcs from $R$ to $V\setminus R$. Furthermore, there is an FPT algorithm for enumerating all important cuts in a directed graph~\cite{cygan2015parameterized}.

\begin{restatable}{lemma}{LemUsefulChecking}\label{lem:useful_checking}
    For a terminal partition $\mathcal T$ with $\mincut_G(\mathcal T)\leq c$, the following are equivalent:
    \begin{itemize}
        \item[\textnormal{\textbf{(i)}}] $\mathcal T$ is a useful terminal partition of $T$, and
        \item[\textnormal{\textbf{(ii)}}] Every minimum multiway cut $F$ of $\mathcal T$ is a connected multiway cut if some component $T'$ in $\mathcal T$ and important cut $R$ of $(T',T\setminus T')$ satisfy $\partial R\subset F$, $|E(R)|\geq c\phi^{-1}$, and $F\cap E(R)=\emptyset$.
    \end{itemize}
    \vspace{0.5em}
\end{restatable}
\begin{proof}
    Note that {$\textbf{(i)}\to \textbf{(ii)}$} direction is trivial since it is consistent with the definition of useful. 
    We demonstrate the reverse direction using contradiction. 
    
    Let $\mathcal T$ be a terminal partition satisfying \textbf{(ii)}.
    For an arbitrary minimum multiway cut $F$ of $\mathcal T$, we show that it is a connected multiway cut which implies \textbf{(i)}.
    If the core of $F$ is $V$, then it is trivial since $G$ is a connected hypergraph.
    Otherwise, the core of $F$ is $V\setminus X$ for a connected component $X$ in $G\setminus F$ with $|E(X)|>c\phi^{-1}$ because $G$ is a $\phi$-expander.
    Recall that a core of a multiway cut is the whole vertex set or the complement set of the largest connected component according to the multiway cut.
    In this proof, we obtain a component $T'$ of $\mathcal T$ and an important cut $\partial R$ with $R\supset X$ outlined in this lemma so that they would be a witness that $T\setminus X$ is connected in $\hat G[V\setminus X]$. 

    Let $T'$ be the terminal component in $\mathcal T$ intersecting $X$.
    If $X$ has no terminal in $T$, then we choose an arbitrary component $T'$ in $\mathcal T$.
    Let $\partial R$ be an important cut of $(\bar X, T\setminus T')$ with $|\partial R|\leq |\partial \bar X|$, where $\bar X$ is the union of all components according to $G\setminus F$ which intersects $T'$.
    We obtain a multiway cut $F'$ of $\mathcal T$ from $F$ by replacing the hyperedges contained in $E(R)$ with $\partial R$. 
    Then the following conditions hold:
    \begin{itemize}
        \item $X\subset R$, $\partial R\subset F$, and $F'\cap E(R)=\emptyset$,
        \item $|E(R)|\geq c\phi^{-1}$, and
        \item $F'$ is a minimum multiway cut of $\mathcal T$.
        \vspace{-0.5em}
    \end{itemize}
    
    The first condition holds by construction.
    The second one holds due to $X\subset R$ and $|E(X)|> c\phi^{-1}$.
    Note that the obtained $F'$ is a multiway cut of $\mathcal T$ by construction.
    Furthermore, its size is at most $|F|-|\partial \bar X|+|\partial R|$ which is at most $|F|$.
    Thus, the last condition also holds. Recall that $F$ is a minimum multiway cut of $\mathcal T$.

    By the above conditions, we have that $F'$ is a minimum multiway cut of $\mathcal T$, and also a connected multiway cut where $V\setminus R$ is the core of $F'$ by \textbf{(ii)}.
    Furthermore, $T\setminus X$ is connected in $\hat G[V\setminus X]$ as it is connected in $\hat G[V\setminus R]$ and $V\setminus R\subset V\subset X$.
    In conclusion, $F$ is a connected multiway cut, and thus, the lemma holds.
\end{proof}

We construct an auxiliary directed graph $D^{\textsf{inc}}$ to enumerate important cuts in $G$.
For instance $(G, T,c)$, the vertex set of $D^{\textsf{inc}}$ is the union of $V(G)$ and two copies $E_{\textsf{in}}$ and $E_{\textsf{out}}$ of $E(G)$. 
For a hyperedge $e$ in $E(G)$, we use $e_{\text{in}}$ and $e_{\text{out}}$ to denote the copies of $e$ in $E_{\textsf{in}}$ and $E_{\textsf{out}}$, respectively, and we insert one arc from $e_{\textsf{in}}$ to $e_{\textsf{out}}$.
For each $v\in V(G)$ and $e\in E(G)$ with $v\in e$, we insert $2c$ parallel arcs from $v$ to $e_{\textsf{in}}$ and from $e_{\textsf{out}}$ to $v$.
Important cuts in $G$ correspond one-to-one with those in $D^{\textsf{inc}}$.

We can enumerate all important cuts of $G$ by applying the FPT algorithm for $D^{\textsf{inc}}$.
There are at most $2^{O(rc)}$ number of important cuts in $G$, and they can be enumerated in $2^{O(rc)}m$ time~\cite{cygan2015parameterized}, where $m=|E(G)|$.
By Lemma~\ref{lem:useful_checking}, we can prune all useful terminal partitions in $|T|(rc\phi^{-1})^{O(rc)}m$ time among the $|T|(rc\phi^{-1})^{O(rc)}$ enumerated candidates.

Lemma~\ref{lem:construct_pag} summarizes this section.
In the remainder, we give an algorithm to compute a minimal multicut-mimicking network for $\phi$-expander using it.
\begin{restatable}{lemma}{ConstructPAG}\label{lem:construct_pag}
    There are $|T|(rc\phi^{-1})^{O(rc)}$ useful partitions and their minimum multiway cuts in a $\phi$-expander $(G,T,c)$.
    We can enumerate all of them in $|T|(rc\phi^{-1})^{O(rc)}m$ time.
\end{restatable}
\begin{proof}
    Recall that  $n$ and $m$ denote the number of vertices and hyperedges, respectively, in $G$.
    We can enumerate all $k(rc\phi^{-1})^{O(rc)}$ number of connected multiway cuts of size at most $c$ in $k(rc\phi^{-1})^{O(rc)}$ time.
    Thus, $k(rc\phi^{-1})^{O(rc)}$ terminal partitions are obtained including all useful partitions. 
    In the following, we select out all useful partitions among them in $k(rc\phi^{-1})^{O(rc)}m$ time. Subsequently, we prune all minimum multiway cuts.

    To prune useful partitions,  we construct a directed graph $D^{\textsf{inc}}$ which preserves all important cuts in $G$ of size at most $c$ in Section~\ref{sec:expander}. 
    The vertex set of $D^{\textsf{inc}}$ is the union of $V(G)$ and two copies $E_{\textsf{in}}$ and $E_{\textsf{out}}$ of $E(G)$. 
    For a hyperedge $e$ in $E(G)$, we use $e_{\text{in}}$ and $e_{\text{out}}$ to denote the copied $e$ in $E_{\textsf{in}}$ and $E_{\textsf{out}}$, respectively.
    For each $v\in V(G)$ and $e\in E(G)$ with $v\in e$, we insert $2c$ parallel arcs from $v$ to $e_{\textsf{in}}$ and from $e_{\textsf{out}}$ to $v$.
    Additionally, we insert one arc from $e_{\textsf{in}}$ to $e_{\textsf{out}}$ for every $e\in E(G)$.
    The obtained directed graph $D^{\textsf{inc}}$ preserves all important cuts of size at most $c$ in $G$ by Observation~\ref{obs:importantcuts}.

    \begin{restatable}{observation}{ObsImportantCuts}\label{obs:importantcuts}
    For disjoint $A, B\subset V(G)$ and $R\supset A$ with $|\partial_G R|\leq c$, $\partial_G R$ is an important cut of $(A,B)$ in $G$ if and only if $\{e_{\textsf {in}}\to e_{\textsf{out}}\mid e\in \partial_G R\}$ is that in $D^{\textsf{inc}}$.
\end{restatable}
\begin{proof}   
    Let $F=\{e_{\textsf {in}}\to e_{\textsf{out}}\mid e\in \partial_G R\}$, and $R_A$ be the reachable vertices from $R$ in $D^{\textsf{inc}}\setminus F$.
    The goal is to show that $\partial R$ is an important cut of $(A, B)$ in $G$ if and only if $\partial_{D^{\textsf{inc}}} R_A$ is that in $D^{\textsf{inc}}$. 
    It is trivial that the vertex set $R_A\cap V(G)$ in $D^{\textsf{inc}}$ contains $R$ in $V(G)$.
    Additionally, the others $V(G)\setminus R$ is excluded by $R_A$.
    The `if' direction naturally holds from this observation.
    Precisely, if $\partial_G R$ is not an important cut in $G$, there is a cut $\{e_{\textsf {in}}\to e_{\textsf{out}}\mid e\in \partial_G R'\}$ in $D^{\textsf{inc}}$ with $R\subsetneq R'$ and $|\partial R|\geq |\partial R'|$.
    The reachable vertex set $R'_A$ from $R'$ in $D^{\textsf{inc}}\setminus F$ is a witness that $\partial_{D^{\textsf{inc}}} R_A$ is not an important cut in $D^{\textsf{inc}}$.
    
    For the `only if ' direction, we suppose that $F$ is not an important cut in $D^{\textsf{inc}}$, and consider $R_A'\supsetneq R_A$ with $|\partial_{D^{\textsf{inc}}}R_A'|$ is at most $|F|$.
    Note that there are 2c number of outgoing arcs  from $v$ to $e_{\textsf{in}}'$ (and from $e_{\textsf{out}}'$ to $v$) in $D^{\textsf{inc}}$ when $v\in e'$.
    Thus, $R_A'$ includes $e_{\textsf{in}}'$ where $e'$ is incident to some vertex in $R_A'\cap V(G)$ in $G$.
    Additionally, $e_{\textsf{out}}'$ is excluded by $R_A'$ where $e'$ is incident to some vertex in $V(G)\setminus R_A'$ in $G$.
    From these observations, we have that $R_A'$ is a proper superset of $R_A$ if and only if $R_A'\cap V(G)$ is a proper superset of $R_A\cap V(G)=R$.
    Furthermore, $|\partial_{D^{\textsf{inc}}} R'_A|\geq |\partial_G (R'_A\cap V(G))|$.
    Thus, $R'_A\cap V(G)$ is the witness that $\partial_G R$ is not an important cut in $G$.
    In conclusion, $\{e_{\textsf {in}}\to e_{\textsf{out}}\mid e\in \partial_G R\}$ is an important cut in $D^{\textsf{inc}}$ if $\partial_G R$ is that in $G$.
\end{proof}

 Therefore, we can obtain all important cuts of size at most $c$ in $G$ by applying the following lemma for $D^{\textsf{inc}}$. 
    
    \begin{lemma}[\protect{\cite[Theorem 8.36]{cygan2015parameterized}}]\label{lem:enumerating_important_cuts}
        For a directed graph $D$ and two disjoint vertex sets $A, B\subset V(D)$, there are $2^{O(c)}$ number of important $(A, B)$-cuts of size at most $c$. Furthermore, all of them can be enumerated in $2^{O(c)}(|E(D)|+|V(D)|)$ time.
    \end{lemma}
    

    Since the directed graph $D^{\textsf{inc}}$ has at most $2rc(n+m)$ arcs, the algorithm outlined in Lemma~\ref{lem:enumerating_important_cuts} enumerates $2^{O(c)}$ important cuts in $G$ in $r^2 2^{O(c)} m$ time. Note that $n\leq rm$.

    For a terminal partition $\mathcal T$ and an important cut $\partial R$ of $(T',T\setminus T')$ where $T'$ is a component of $\mathcal T$,
    we can check the condition \textbf{(ii)} of Lemma~\ref{lem:useful_checking} in $(c\phi^{-1})^{O(c)} m$ time.
    Precisely, we first check whether $|E(R)|\geq c\phi^{-1}$ and $|\partial R|\leq c$. 
    Then check all $(c\phi^{-1})^{O(c)}$ number of minimum multiway cuts $F$ are connected multiway cut or not where $F$ is the union of $\partial R$ and a subset of $E(V\setminus R)$ of size at most $c$.
    Note that $|E(V\setminus R)|$ is at most $c\phi^{-1}$ since our instance is a $\phi$-expander.
    
    Since there are at most $rc$ components in $\mathcal T$, we can check whether a terminal partition $\mathcal T$ is a useful terminal partition or not in $r^2(c\phi^{-1})^{O(c)} m$ time by Lemma~\ref{lem:useful_essential} and Lemma~\ref{lem:enumerating_important_cuts}.
    In conclusion, we can prune all useful partition in $k(rc\phi^{-1})^{O(rc)} m$ time among $k(rc\phi^{-1})^{O(rc)}$ candidates.
    In the following, we prune all minimum multiway cuts of the useful terminal partitions among $k(rc\phi^{-1})^{O(rc)}$ candidates.

    When we have exactly all useful partitions, we can prune all minimum multiway cuts among the enumerated multiway cuts at the first phase by comparing the sizes of multiway cuts contributing the same useful partition of $T$.
    Recall that a minimum multiway cut of a useful partition is a connected multiway cut.

    In conclusion, we can enumerate all useful terminal partitions and their minimum multiway cuts in $k(rc\phi^{-1})^{O(rc)}m$ time. Furthermore, the number of them is at most $k(rc\phi^{-1})^{O(rc)}$.
\end{proof}

\subsection{Algorithm for $\phi$-Expanders}\label{sec:utilizing_pag}
In this section, we illustrate how to obtain a \emph{minimal} instance $H$ with respect to a $\phi$-expander $(G, T,c)$ for $\phi>0$. 
Precisely, we give a detailed proof for Lemma~\ref{lem:computing_essential_in_pag}. 
Note that a minimal multicut-mimicking network has at most $|T|c^{O(r\log c)}$ hyperedges by Theorem~\ref{thm:exists}.
Here, $k=|T|$, $m=|E(G)|$, and $r$ is the rank of $G$.
\begin{restatable}{lemma}{LemEssentialPag}\label{lem:computing_essential_in_pag}
    For a $\phi$-expander $(G, T,c)$, we can find a minimal multicut-mimicking network in $|T|(rc\phi^{-1})^{O(rc)}m$ time. Moreover, it has at most $|T|c^{O(r\log c)}$ hyperedges.
\end{restatable}

\subparagraph*{Algorithm.}If the instance has at most $O(c\phi^{-1})$ hyperedges, then we enumerate all multiway cuts where the size is at most $c$, and their consisting terminal partitions.
Then we find a non-essential hyperedge $e$ and contract it.
Recursively, we find and contract a non-essential hyperedge in the $(G/e, T,c)$ until in which every hyperedge is essential.

If $G$ has more than $(3c\phi^{-1}+c)$ hyperedges, then we detect all essential hyperedges using Lemma~\ref{lem:useful_essential}.
Precisely, we enumerate all useful partitions and their minimum multiway cuts where the size is at most $c$ using Lemma~\ref{lem:construct_pag}. 
Recall that the enumerated multiway cuts are connected multiway cuts.
Then we check whether a hyperedge $e$ is essential by detecting all minimum multiway cuts containing $e$ and all useful partitions contributed by them.
Precisely, for each of the useful partitions, check whether another minimum multiway cut exists except them while excluding $e$.
If $e$ is essential, then we move to the next hyperedge.
Otherwise, we contract $e$ and move to the next hyperedge for detecting non-essential hyperedges in $(G/e, T,c)$.
After the first iteration, we do not call the algorithm again outlined in Lemma~\ref{lem:construct_pag}.
Instead, we delete every multiway cut containing $e$ among the enumerated minimum multiway cuts before moving to the next iteration, where $e$ is the contracted non-essential hyperedge.

We visit every hyperedge at most once until \textbf{(i)} at most $(3c\phi^{-1}+c)$ hyperedges are left or \textbf{(ii)} every hyperedge has been visited.
When we reach the case \textbf{(i)}, we compute a multicut-mimicking network of the current instance using the algorithm for the case that $m\in O(c\phi^{-1})$ which is already illustrated.
In the latter case, every hyperedges is essential in the current instance. Thus, we return it as a solution of the original instance.
\subparagraph*{Correctness.}
Note that a contraction of a $\phi$-expander is also a $\phi$-expander.
Furthermore, contraction of a non-essential hyperedge is a multicut-mimicking network by Lemma~\ref{lem:contracting}.

The algorithm for the instance with $m\leq (3c\phi^{-1}+c)$ is trivial since it is a brute force algorithm.
For the instance with $m>(3c\phi^{-1}+c)$, its correctness is guaranteed by Lemma~\ref{lem:useful_essential} and an observation: no new connected multiway cut is occurred by contracting a non-essential hyperedge while $m>(3c\phi^{-1}+c)$.
Precisely, if $(G/e, T,c)$ has more than $(3c\phi^{-1}+c)$ hyperedges, a connected multiway cut in $(G/e, T,c)$ is also that in $(G, T,c)$.
This observation implies that all useful partitions and their minimum multiway cuts in $(G/e, T,c)$ were already enumerated at the beginning of this algorithm. In the following, we prove the observation.

For a multiway cut $F$ in a contraction $G/e$ with $|F|\leq c$, it corresponds to a multiway cut $\bar F$ in the original graph $G$ with the same size. 
This implies that if $F$ is a minimum multiway cut of a terminal partition in $G/e$, then $\bar F$ is also that in $G$.
Note that $F$ does not contain $e$.
We suppose that $F$ is a connected multiway cut of $(G/e, T,c)$, and we show that $\bar F$ is that in $(G, T,c)$.
Let $C$ and $\bar C$ be the cores of $F$ in $G/e$ and $\bar F$ in $G$, respectively.
If $C$ and $\bar C$ are equal, then $\bar F$ is a connected multiway cut.
For the other case, there are two possible cases: $C\subset \bar C$ or $\bar C\subset C$.
This is because the size of a connected component in $G\setminus \bar F$ cannot be increased by contracting a hyperedge $e$ not in $\bar F$.
The first case $C\subset \bar C$ occurs only if $\bar C$ contains the hyperedge $e$.
Then $T\cap C=T\cap \bar C$ is connected in $\hat G[\bar C]$, and thus, $\bar F$ is a connected multiway cut in $G$.
The latter case $\bar C\subset C$ only if the size of $V\setminus C$ is decreased under $c\phi^{-1}$ by contracting $e$.
In such a case, $C=V(G/e)$. That means $|E(G/e)|=|E(C)|$ and it is at most $3c\phi^{-1}+c$ by Observation~\ref{obs:property_connected_expander}.
\subparagraph*{Time complexity.}
 Recall that if $G$ has $O(c\phi^{-1})$ hyperedges, there are $(c\phi^{-1})^{O(c)}$ number of multiway cuts of size at most $c$. Moreover, at most $(rc\phi^{-1})^{O(rc)}$ number of terminal partitions are obtained by the multiway cuts.
    Then we can find all essential hyperedges in $(rc\phi^{-1})^{O(rc)}$ time, and also a non-essential hyperedge $e$ if exists.
    Therefore, we can meet a point that all hyperedges are essential before $O(c\phi^{-1})$ iterations, it takes $(rc\phi^{-1})^{O(rc)}$ time. In the following, we analyze the case for $m>(3c\phi^{-1}+c)$.

    We enumerate all useful partitions and their minimum multiway cuts using Lemma~\ref{lem:construct_pag}.
    It takes $k(rc\phi^{-1})^{O(rc)}m$ time and $k(rc\phi^{-1})^{O(rc)}$ number of connected multiway cuts and useful partitions are given. 
    Moreover, each multiway cut constitutes at most $(rc)^{O(rc)}$ number of useful partitions.
    Thus, determining a hyperedge $e$ is essential takes $|\mathcal F_e|(rc)^{O(rc)}$ time, where $|\mathcal F_e|$ denotes the number of connected multiway cuts including $e$.
    In the same time complexity, we can delete all connected multiway cuts containing $e$.
    Since each multiway cut consists of at most $c$ hyperedges, $\sum |\mathcal F_e|$ is at most $k(rc\phi^{-1})^{O(rc)}$.
    Totally, the algorithm takes $k(rc\phi^{-1})^{O(rc)}m$ time until we meet points \textbf{(i)} or \textbf{(ii)}. When we halt at \textbf{(ii)} then the algorithm returns the current instance, and at \textbf{(i)} it requires $(rc\phi^{-1})^{O(rc)}$ additional time.
    In conclusion, the algorithm takes $k(rc\phi^{-1})^{O(rc)}m$ time.



\subsection{Near-Linear Time Algorithm for General Hypergraphs}\label{sec:algorihtm_near_linear}


In this section, we obtain a multicut-mimicking network for a general instance $(G, T,c)$, by recursively calling \textsf{MimickingExpander}.
The submodule \textsf{MimickingExpander} computes a small multicut-mimicking network based on
the \emph{expander decomposition} of Long and Saranurak~\cite{long2022near} and
the algorithm for a $\phi$-expander with $\phi > 0$ outlined in Lemma~\ref{lem:computing_essential_in_pag}.
However, its return is not sufficiently small, and thus, we obtain a much smaller solution by applying it recursively.



\subparagraph*{{MimickingExpander$(G;T,c)$}.}
We let $n=|V(G)|$, $m=|E(G)|$, and $\phi^{-1}=4rc^{Mr\log c}\log^3 n$, where the multicut-mimicking network returned by Lemma~\ref{lem:computing_essential_in_pag} has at most $kc^{Mr\log c}$ hyperedges for an expander with $k$ terminals.
When the submodule is called, we first decompose the vertex set $V(G)$ into the vertex partition $(V_1, \dots, V_s)$ 
so that the size of the cut of $(V_1,\dots, V_s)$ is at most $\phi m \log ^3 n$ and each $\hat G[V_j]$ is a $\phi$-expander for $j\in[s]$.
There is a $p^{1+o(1)}$ time algorithm computing such a vertex partition~\cite{long2022near}, where $p=\sum_{e\in E}|e|$. 
\begin{lemma}[\protect{\cite[Theorem C.3]{long2022near}}]\label{lem:decompose_expander}
    For a hypergraph $G$ and $\phi^{-1}\in m^{o(1)}$,
    There exists a randomized algorithm computing a vertex partition $(V_1,\dots,V_s)$ in $p^{1+o(1)}$ time  so that:
    \begin{itemize}
        \item {$\hat G[V_j]$ is a $\phi$-expander for $j\in[s]$ and}
        \item{the size of $(V_1,\dots,V_s)$ cut in $G$ is at most $\phi m \log^3 n$,}
    \end{itemize}
     where $\hat G[V_i]$ is the separated subgraph of $G$ with respect to $V_i$ for $i\in[s]$ and $p=\sum_{e\in E}|e|$.
    
\end{lemma}

Every subinstance $(\hat G[V_j], T_j, c_j)$ is a  $\phi$-expander, where $c_j=\min\{c,|T_j|\}$ and $T_j$ is the union of $T \cap V_j$ and anchored terminals in $\hat G[V_j]$. 
For each $(\hat G[V_j],T_j,c_j)$, we construct a multicut-mimicking  network $H_j$ by Lemma~\ref{lem:computing_essential_in_pag}.
Finally, we return the multicut-mimicking network $H$ by gluing $H_1,\dots, H_s$. 
Precisely, we merge all restricted hyperedges having a common anchored terminal and remove all anchored terminals not in $V$.
The following lemma summarizes the performance of this submodule.

\begin{restatable}{lemma}{LemSubmodule}\label{lem:submodule_near_linear}
    \textup{\textsf{MimickingExpander$(G;T,c)$}} returns a multicut-mimicking  network of $(G,T,c)$ with $(|T|c^{O(r\log c)}+(m/2))$ hyperedges in $(p^{1+o(1)}+|T|(c^{r\log c}\log n)^{O(rc)} m)$ time.
    \vspace{-0.5em}
\end{restatable}
\begin{proof}
Recall that $\phi^{-1}=4rc^{Mr\log c}\log^3 n$.
The time complexity of this algorithm holds by Lemma~\ref{lem:computing_essential_in_pag} and Lemma~\ref{lem:decompose_expander}.
We first demonstrate that the returned $H$ has at most $|T|c^{O(r\log c)}+m/2$ hyperedges.
Note that each $H_j$ has at most $|T_j|c^{Mr\log c}$ hyperedges.
Additionally, the size of the cut of $(V_1, \dots, V_s)$ is limited to $\phi m\log^3 n$ by Lemma~\ref{lem:decompose_expander}, and thus, we have that $\sum_{j\in [s]} |T_j|$ is at most $(k+2r\phi m\log^3 n)$. 
Consequently, the returned $H$ has at most $(k+2r\phi m\log^3 n)c^{Mr\log c}$ hyperedges.
Since we set $\phi^{-1}=4rc^{Mr\log c}\log^3 n$, $H$ has at most $|T|c^{O(r\log c)}+m/2$ hyperedges.
In the following, we demonstrate that the obtained $H$ from gluing $H_1,\ldots, H_s$ is a multicut-mimicking network of $(G, T,c)$. 

Recall that the algorithm outlined in Lemma~\ref{lem:computing_essential_in_pag} obtains $H_j$ by contracting non-essential hyperedges $F_j$ in $(\hat G[V_j], T_j, c_j)$ sequentially for $j\in [s]$.
We define an ordering $e\prec_j e'$ between two contracted non-essential hyperedges in $F_j$ when $e$ is contracted before $e'$ in the algorithm.
Furthermore, we extend the orderings $\prec_j$'s as $\preceq$ for every contracted hyperedges $\cup_{j\in [s]} F_j$ so that $e\prec e'$ if $e\prec_j e'$ for some $j\in [s]$, otherwise, $e\preceq e'$ and $e'\preceq e$. Then $H$ is the same as the contraction in $G$ by contracting $\cup F_j$ sequentially in increasing order of $\preceq$.
This is because the cut of $(V_1,\dots,V_s)$ are not contracted.

We fix a hyperedge $e\in F_j$.
By Lemma~\ref{lem:essential_in_subinstance} and Lemma~\ref{lem:computing_essential_in_pag}, $e$ is non-essential in $(G',T,c)$, where $G'$ is obtained by contracting all hyperedges $e'\preceq e$ in $G$ sequentially.
Precisely, $e$ is non-essential in the contraction of all $e'\in F_j$ with $e'\prec_j e$ in $(\hat G[V_j],T_j, c_j)$ by the algorithm outlined in Lemma~\ref{lem:computing_essential_in_pag}.
Moreover, the contraction constitutes a subinstance of $(G', T,c)$ since the other hyperedges contracted before $e$ do not intersect $V_j$. Thus, $e$ is a non-essential hyperedge in $(G',T,c)$ by Lemma~\ref{lem:essential_in_subinstance}. 
In conclusion, $H$ is a contraction in $G$ and a multicut-mimicking network of $(G, T,c)$.
\end{proof}

\subparagraph*{Overall algorithm.}For an instance $(G,T,c)$, we initialize $G_0 = G$ and obtain the multicut-mimicking network $G_i$ by \textsf{MimickingExpander$(G_{i-1}; T, c)$} for $i\in[\lceil \log m\rceil]$, inductively. Finally, we return $G_{\lceil \log m\rceil}$.
This algorithm corresponds to Theorem~\ref{thm:near_linear_algo}.
\thmNearLinear*
\begin{proof}
Every $G_{i}$ for $i\in[\lceil \log m \rceil]$ is a multicut-mimicking  network for the original instance $(G, T, c)$ by Lemma~\ref{lem:submodule_near_linear}. 
We show that $G_{\lceil \log m\rceil}$ has $kc^{O(\log c)}$ hyperedges.

For $i\in [\lceil \log m\rceil]$, let $m_i$ denote the number of hyperedges in $G_{i}$ for $i \in [\lceil \log m\rceil]$.
Then the recurrence relation $m_i\leq ( kc^{O(r\log c)}+\frac{m_{i-1}}2)$ holds by Lemma~\ref{lem:submodule_near_linear} along with $m_0=m$.
Inductively, we can bound $m_{\lceil \log m \rceil}$, which is the number of hyperedges in the returned multicut-mimicking network $G_{\lceil \log m\rceil}$, as $kc^{O(r\log c)}$.

In conclusion, we return a multicut-mimicking network of $(G, T,c)$ with at most  $kc^{O(r\log c)}$ hyperedges. 
Furthermore,
calling \textsf{MimickingExpander}'s takes $p^{1+o(1)}+k(c^{r\log c}\log n)^{O(rc)}\cdot(m_0+m_1+\dots+m_{\lceil \log m\rceil-1})$ time by Lemma~\ref{lem:submodule_near_linear}.
Furthermore, $(m_0+\dots+m_{\lceil \log m\rceil-1})$ is at most $(2m+kc^{O(r\log c)}\lceil\log m\rceil)$.
For the sufficiently large $m$ with $kc^{O(r\log c)}\in m^{o(1)}$, the time complexity is in $(p^{1+o(1)}+k(c^{r\log c}\log n)^{O(rc)}\cdot m)$ time. The proof is complete.
\end{proof}

\section{Bound for Minimal Instances}\label{sec:exists}
To complete the proof of Lemma~\ref{lem:computing_essential_in_pag} (and Theorem~\ref{thm:near_linear_algo}), we need to show that a minimal multicut-mimicking network for an expander $(G, T,c)$ has at most $|T|c^{O(r\log c)}$ hyperedges.
This section demonstrates it for not only expanders but also general instances.
Precisely, we show the following theorem in this section. 
Here, $r$ is the rank of $G$.
\begin{restatable}{theorem}{thmExists}\label{thm:exists}
    Every minimal instance $(G,T,c)$ has at most $|T|c^{O(r\log c)}$ hyperedges.
\end{restatable}
In this section, we consider the scenario that every terminal in $T$ has degree one in $G$. 
It is sufficient since the other case can be reduced to this scenario by inserting $c+1$ dummy terminals instead of each terminal 
$t$ in $T$ that are adjacent only to $t$. 
This reduction does not increase the rank or the parameter $c$ while increasing the number of terminals by at most $c$ times. However, this increase does not affect the asymptotic complexity in Theorem~\ref{thm:exists}. 
The following explains the previous works and introduces the notions used in this section.

We broadly follow the approach of Wahlstr{\"o}m~\cite{wahlstrom2022quasipolynomial_multicut}.
He utilized the framework of Kratsch et al.~\cite{kratsch2012representative} for multicut-mimicking networks in graphs without the parameter $c$. 
We incorporate the \emph{unbreakable} concept to address the parameter $c$. 
Additionally, we slightly modify the \emph{dense} concept used in the previous work to account for hypergraphs.
\subparagraph*{Unbreakable and dense.}
For a subinstance $(\hat G[X], T_X, c_X)$ with respect to $X \subset V(G)$, the terminal set $T_X$ includes $T \cap X$ and up to two anchored terminals for each restricted hyperedge $e \in \partial_G X$. Hence, $|T_X| \leq 2|\partial_G X| + |T \cap X|$.
We denote the value $2|\partial_G X| + |T \cap X|$ as $\capacity_T(X; G)$, or $\capacity_T(X)$ if the context is clear. Furthermore, we define the following:
\begin{itemize}
    \item \textbf{Unbreakable:} An instance $(G,T,c)$ is said to be $d$-unbreakable for $d>0$ if 
    $|T\cap X|\leq d$
    for any vertex set $X\subseteq V(G)$ with $|T\cap X|\leq |T\setminus X|$ and $|\partial X|\leq c$.
    \item \textbf{Dense:} An instance $(G,T,c)$ is said to be $\alpha$-dense for $\alpha>0$ if 
    $ |E(X)|\leq (\capacity_T(X))^{\alpha} $ for any vertex set $X\subseteq V$ with $0<|E(X)|\leq |E(V\setminus X)|$ and $|\partial X|\leq c$.
    \vspace{-0.3em}
\end{itemize}

Wahlstr{\"o}m~\cite{wahlstrom2022quasipolynomial_multicut} also used the concept of \emph{dense} defining it based on the vertices instead of $E(\cdot)$. Since he addressed graphs, it was guaranteed that $|E(X)| = \Omega(|X|)$ by using connective assumption.
However, this is not for hypergraphs. Thus, we slightly modify the definition.

In Section~\ref{sec:exists_base}, we prove Theorem~\ref{thm:exists} for \emph{unbreakable and dense} instances using the notions introduced in Section~\ref{sec:matroid}.
Section~\ref{sec:exi_unbr_sparse} explains how to extend this proof for general instances.

\subsection{Matroids and Representative Sets}\label{sec:matroid}
We use the notion of matroids and representative sets as in the previous work, which is a generalization of the notion of linear independence in vector spaces. Formally, a matroid $(S, \mathcal{I})$ consists of a \emph{universe set} $S$ and an \emph{independent set} $\mathcal{I} \subseteq 2^S$ with $\emptyset \in \mathcal I$ satisfying: 
\begin{itemize}
    \item {If $B \in \mathcal{I}$ and $A \subseteq B$, then $A \in \mathcal{I}$, and}
    \item {If $A, B \in \mathcal{I}$ with $|A| < |B|$, then there exists $x \in B \setminus A$ such that $A \cup \{x\} \in \mathcal{I}$.}
\end{itemize}
\vspace{-0.5em}

For a matroid $(S, \mathcal{I})$, its \emph{rank} is the size of the largest set in $\mathcal I$.
It is said to be \emph{representable} if there is a matrix over a field whose columns are indexed by the elements of $S$ such that: $F \subset S$ is in $\mathcal{I}$ if and only if the columns indexed by $F$ are linearly independent over the field.

\subparagraph{Representative sets.} Kratsch et al.~\cite{kratsch2012representative} introduced a framework for computing non-essential vertices using the notion of \emph{representative sets}.
We employ the framework.
For this purpose, we introduce two operations: \emph{truncation} and \emph{direct sum} along with Lemma~\ref{lem:representative_set}.
For a matroid $(S,\mathcal I)$ and an integer $r > 0$, the $r$-\emph{truncation} of $(S,\mathcal I)$ is defined as a matroid $(S, \mathcal I')$ such that $F\subseteq S$ is contained in $\mathcal I'$ if and only if $|F|\leq r$ and $F\in \mathcal I$. 
Note that an $r$-truncation has rank at most $r$.
For matroids $\mathcal{M}_1, \dots, \mathcal{M}_s$ over disjoint universes with $\mathcal M_i=(S_i, \mathcal I_i)$ for $i\in [s]$, 
their \emph{direct sum} is defined as a matroid $(S,\mathcal I)$ such that $S$ is the union of all $S_i$, and a subset $F$ of $S$ is in $\mathcal I$ if and only if $F$ can be decomposed into $s$ disjoint subsets, each of which is independent in $\mathcal{M}_i$.
The direct sum of matroids also satisfies the matroid axioms. 

For a matroid $(S, \mathcal{I})$ and two subsets $A,B$ of $S$, 
we say $A$ \emph{extends} $B$ if $A \cap B = \emptyset$ and $A \cup B \in \mathcal{I}$.
For $\mathcal{J} \subseteq 2^{S}$, a subset $\mathcal{J}^*$ of $\mathcal{J}$ is called 
a \emph{representative set} if for any set $B \subseteq S$, there is a set $A^* \in \mathcal{J}^*$ that extends $B$ when there is a set $A \in \mathcal{J}$ that extends $B$. 
\vspace{-0.5em}
\begin{lemma}[\protect{\cite[Lemma 3.4]{kratsch2012representative}}]\label{lem:representative_set}
    Let $\mathcal{M}_1 = (S_1, \mathcal{I}_1), \dots, \mathcal{M}_s = (S_s, \mathcal{I}_s)$ be matroids represented over the same finite field and with pairwise disjoint universe sets. 
    Let $\mathcal{J}$ be a collection of sets containing one element from $S_i$ for each $i \in [s]$. 
    Then, some representative set of $\mathcal J$ in the direct sum of all matroids $\mathcal{M}_i$ has a size at most the product of the ranks of all $\mathcal M_i$.
\end{lemma}
\vspace{-0.5em}
\subparagraph{Uniform and (hyperedge) gammoids.}
We use two classes of representable matroids: \emph{uniform matroids} and \emph{gammoids}. 
They, along with their truncation and direct sum, are representable over any large field~\cite{cygan2015parameterized,kratsch2012representative,lokshtanov2018deterministic,marx2009parameterized,oxley2022matroid}.
\vspace{-0.5em}

\begin{itemize}
    \item \textbf{Uniform matroid:} For a set $S$ and an integer $r > 0$, the \emph{uniform matroid} on $S$ of rank $r$ is the matroid in which the universe set is $S$ and the independent set consists of the sets containing at most $r$ elements in $S$. 
    \item \textbf{Gammoid:} 
    For a directed graph $D=(V,A)$ and two subsets $S$ and $U$ of $V$, 
    a \emph{gammoid} defined on $(D,S,U)$
    is a matroid $(U, \mathcal I)$ where $\mathcal I$ 
    consists of the sets $X\subseteq U$ such that there are $|X|$ pairwise vertex-disjoint paths in $D$ from $S$ to $X$.
\end{itemize}
\vspace{-0.5em}

In this paper, we deal with \emph{hyperedge} multicuts/multiway cuts in hypergraphs, and thus we need the hypergraphic counterparts. 
Therefore, we define and use 
\emph{hyperedge gammoids}.

\begin{itemize}
    \item \textbf{Hyperedge gammoid:} 
    For a hypergraph $G$ and a terminal set $T\subset V(G)$, we consider a directed graph $D^{\textsf{split}}$ defined as follows. 
    First, we start from the undirected graph $D^{\textsf{split}}$ of which the vertex set is $E(G)$ and $ee'\in E(D^{\textsf{split}})$ for two $e,e'$ in $E(G)$ (and $V(D^{\textsf{split}})$) if and only if $e\cap e'$ is not empty.
    Then, we replace each undirected edge with two-way directed arcs.
    Furthermore, we insert a copy $E_{\sink}$ of $E(G)$ into $V(D^{\textsf{split}})$.
    We call the copy in $E_\sink$ of a hyperedge $e\in E(G)$ a \emph{sink-only copy} of $e$, and denote it by \textsf{sink}$(e)$. We insert one-way arcs from $e'$ to $\sink(e)$ on $D^{\textsf{split}}$ for every $e'\in E(G)$ where $e'\cap e$ is not empty.
The \emph{hyperedge gammoid} of $(G,T)$ is the gammoid on $(D^{\textsf{split}},\partial_G T\subset E(G),E(G)\cup E_\sink)$.  
\end{itemize}

Recall that a \emph{path} of a hypergraph is defined as a sequence of hyperedges such that any two consecutive hyperedges contain a common vertex. 
If a path starts or ends at a hyperedge containing a terminal $t$, we say that it starts or ends at $t$ to make the description easier. 
For a subset $F$ of $E(G)\cup E_\sink$, let $F_E$ be the set of hyperedges $e$ of $E(G)$ where $e$ or $\sink(e)$ is contained in $F$.
Even if $F$ contains both $e$ and $\sink(e)$, $e$ appears in $F_E$ exactly once. 

\begin{restatable}{observation}{ObservationHyperGammoid}\label{obs:hypergammoid}
$F\subseteq E(G)\cup E_\sink$ is independent in the hyperedge gammoid of $(G,T)$ if and only if 
there exist $|F|$ pairwise edge-disjoint paths from $T$ to $F_E$ in $G$ with two exceptions:
\begin{itemize}
    \item Two paths can end at $e\in E(G)$ if $e\in F$ and $\sink(e) \in F$, and
    \item One path can pass through $e$ while another path ends at $e$ if  $e\notin F$ but $\sink(e)\in F$.
\end{itemize}
\end{restatable}
\begin{proof}
    By the construction, $F$ is independent in the hyperedge gammoid of $(G, T,c)$ if and only if there exist $|F|$ pairwise vertex-disjoint paths in $D^{\textsf{split}}$ from $\partial T$ to $F$. 
    These vertex-disjoint paths in $D^{\textsf{split}}$ correspond to edge-disjoint paths in $G$, except at the hyperedges and their sink-only copies. 
    Since sink-only copies have no outgoing arcs, they do not appear at the starting or internal nodes of the paths. Thus, the observation holds.
\end{proof}

\subsection{Essential Hyperedges in Unbreakable and Dense Instances}\label{sec:exists_base}
Assume that $(G, T, c)$ is $d$-unbreakable and $\alpha$-dense with $c\leq d\leq |T|$ and $\alpha\geq 35r\log d$, where $r$ is the rank of $G$.
In this section, we show that $(G, T,c)$ contains at most $|T|d^{\alpha-1}$ essential hyperedges which directly implies Theorem~\ref{thm:exists} for $O(c)$-unbreakable and $\Theta (r\log c)$-dense instances.
Precisely, if there are more than $|T|d^{\alpha-1}$ hyperedges, then at least one is non-essential, and thus, the instance is not minimal.
Here, $k=|T|$ and $r$ is the rank of $G$.

The following matroids would be a witness for our claim with $i_0=30r$:
    \begin{itemize}
        \item {One uniform matroid on {$E$} of rank $(\kappa +c)$, where $\kappa=({d}/2)^{\alpha-i_0-2}$,}
        \item {One $(k+c+1)$-truncation $\mathcal M_0$ of the hyperedge gammoid of $(G,T)$, and}
        \item{$i_0$ copies $\mathcal M_1,\dots, \mathcal M_{i_0}$ of the $(d+c+1)$-truncation of the hyperedge gammoid of $(G,T)$,}
    \end{itemize}
    
    For a hyperedge $e$, its appearance in the universe sets of the uniform matroid on $E$ is denoted by  $e^{\textsf u}$.
    In the universe set of $\mathcal M_i$ for $i\in[0, i_0]$, $e_i$ and $\textsf{sink}_i(e)$ denote the hyperedge and its sink-only copy, respectively. 
    Note that sink-only copies have no outgoing arc in $D^{\textsf{split}}$, where the hyperedge gammoid is defined on the directed graph $D^{\textsf{split}}$, refer to Section~\ref{sec:matroid}.

    Let $\mathcal M$ denote the direct sum of the $(i_0+2)$ matroids described above. 
    For a hyperedge $e\in E(G)$, let $J(e)=\{e^{\textsf u}\}\cup \{\sink_0(e),\dots,\sink_{i_0}(e)\}$. 
    Then, we let $\mathcal J^*$ be the representative set of
    $\{J(e)\mid e\in E(G)\}$ outlined in Lemma~\ref{lem:representative_set}. 
    The set $\mathcal J^*$ consists of all essential hyperedges.
    

    \begin{lemma}\label{lem:existence_corr}
        $\mathcal J^*$ contains all $J(\bar e)$ where $\bar e$ is an essential hyperedge for $(G,T,c)$.
    \end{lemma}
    \begin{proof}
        We fix an essential hyperedge $\bar e$ and show that $J(\bar e)$ is in $\mathcal J^*$.
        To do this, we construct an independent set in $\mathcal M$ extended by  $J(\bar e)$, but not extended by $J( e)$ for any hyperedge $e\neq \bar e$. 
        Let $\mathcal{T}$ be a terminal partition with 
        $\mincut_G(\mathcal{T}) \leq c$ such that every minimum multiway cut of $\mathcal T$ includes   $\bar{e}$. 
        We fix a minimum multiway cut $F$ of $\mathcal T$ in $G$, and let $(V_0, \dots, V_s)$ be the vertex partition according to $G\setminus F$.
        We assume that the component $V_0$ maximizes $|T\cap V_0|$ among $V_0,\dots, V_s$, and $V_1$ maximizes $|E(V_1)|$ among $V_1,\dots, V_s$.
        Additionally, the other components $V_2,\dots,V_s$ are sorted in the decreasing order of $\capacity_T(\cdot)$ values.
        Let $E_\textsf{small}$ denote the union of $E(V_{i_0+1}), E(V_{i_0+2}),\ldots, E(V_s)$.
        
        Then we consider the following sets whose union will be a witness showing that $J(\bar e)$ is in any representative set of $\{J(e)\mid e\in E(G)\}$.
        Let $A^{\textsf u}$ denote the subset $\{ e^{\textsf u} \mid e\in (E_{\textsf{small}}\cup F)\setminus \{\bar e\} \}$ of the universe set of the uniform matroid on $E$ of rank $\kappa+c$. 
        Let $A_i^{\textsf g}$ be the subset $\{ e_i \mid e\in F\cup \partial(T\cap V_i)\}$ of the universe set of $\mathcal M_i$ for $i\in[0,i_0]$.
        We need to demonstrate three assertions: \textbf{(i)} $A^\textsf{u}$ is extended by $\{(\bar e)^{\textsf{u}}\}$ in the uniform matroid on $E$ of rank $\kappa+c$,
        \textbf{(ii)} $A_i^{\textsf{g}}$ is extended by $\{\sink_i(\bar e)\}$ in $\mathcal M_i$ for $i\in [0,i_0]$, 
        and \textbf{(iii)} any $J(e)$ with $e\neq \bar e$ cannot extend $A=A^{\textsf u} \cup  (\cup_{i \in [0, i_0]} A^\textsf{g}_i)$ in $\mathcal M$.
        By combining these assertions, we can conclude that $A$ is extended in $\mathcal M$ by $J(\bar e)$ only, which completes the proof of Lemma~\ref{lem:existence_corr}.

        \subparagraph{Assertion (i).}
        Recall that the size of $F$ is at most $c$. 
        Then $A^\textsf{u}$ is extended by $\{(\bar e)^{\textsf{u}}\}$ in the uniform matroid on $E$ of rank $\kappa+c$ if $|E_{\textsf{small}}|\leq \kappa$ because $\bar e$ is not in $(E_{\textsf{small}}\cup F)\setminus \{\bar e\}$.

        We first show that $\sum_{j\in [s]}\capacity_T(V_j)\leq (3d+2rc)$.
        Note that every component $V_j$ has at most $d$ terminals of $T$ for $j\in[s]$ because $(G,T,c)$ is $d$-unbreakable. 
        Precisely, if a component $V_j$ has more than $d$ terminals, then $V_0$ also contains more than $d$ terminals since we chose $V_0$ as containing the most terminals, and thus, the $d$-unbreakable property is violated from $|V_j\cap T|$, $|T\setminus V_j|>d$.
        Let $x$ be the smallest index in $[s]$ such that $\sum_{j\in[x]}|T\cap V_{j}|>d$.
        By choosing $x$ in this manner, the total size of $T\cap V_{j}$ for all indices $1\leq j<x$ is at most $d$, and for all indices $j>x$ is also at most $d$ due to the $d$-unbreakable property. Due to $|T\cap V_x|\leq d$, we have $\sum_{j\in[s]}|T\cap V_j|\leq 3d$. 
        Moreover, since each hyperedge in $F$ can intersect at most $r$ components and $\capacity_T(V_j)= 2|\partial_G V_j|+|T\cap V_j|$, we have $\sum_{j\in [s]}\capacity_T(V_j)\leq (3d+2rc)$.

            Now we focus on the components $ V_{i_0+1},\ldots, V_s$ as follows. Since $(G,T,c)$ is $\alpha$-dense, $|E(V_j)|\leq (\capacity_T(V_j))^{\alpha}$ for $j\in[2,s]$ by the $\alpha$-dense property.\footnote{
            $|E(V_j)|=\min\{|E(V_j)|,|E(V\setminus V_j)|\}\leq \capacity_T(V_j)^{\alpha}$ for $j\in[2,s]$.}
            Thus, the ordering $\capacity_T(V_2)\geq \dots\geq \capacity_T(V_s)$ implies that $\capacity_T(V_j)\leq (3d+2rc)/(j-1)$. Therefore, we have
        \begin{align*}
            |E_{\textsf{small}}|=\sum_{j\in[i_0+1,s]} |E(V_j)| 
            &\leq \sum_{j\in[i_0+1,s]} \left(\frac{3d+2rc}{i_0}\right)^\alpha
            \leq \sum_{j\in[i_0+1,s]} \left(\frac {d} 4\right)^{\alpha} \\
            &\leq \left({d} /4\right)^{\alpha} \cdot  {rc}
            \leq \left({d} /2\right)^{\alpha+1} \cdot \left(1 /2\right)^{\alpha-1}r
            \leq \left({d} /2\right)^{\alpha-i_0-2}=\kappa.
        \end{align*}
        The inequalities follow from $i_0=30r$, $s \leq rc$, and the assumption we made at the beginning of this subsection: $c\leq d\leq k$ and 
        $\alpha\geq 35r\log d \geq (i_0+2)\log d$.\footnote{Precisely, these assumption give us $r\cdot \left(\frac 1 2\right)^{\alpha-1}\leq \left(\frac d 2\right)^{-i_0-3}$ which implies the last inequality.}      
        It implies {$|A^{\textsf{u}}|\leq \kappa+c-1$} due to $(\bar e)^{\textsf{u}}\notin A^{\textsf u}$.
        Therefore, $\{(\bar e)^{\textsf{u}}\}$ extends $A^{\textsf{u}}$ in the uniform matroid on $E$ of rank $\kappa+c$.
        
         \subparagraph*{Assertion \textsf{(ii)}.}         
        As we observed in the proof of \textbf{Assertion(i)}, each component $V_j$ has at most $d$ terminals for $j\in [s]$. 
        Additionally, it is clear that $|T\cap V_0|\leq |T|=k$.
        These imply that $|A_0^\textsf g|\leq c+k$ and $|A_i^\textsf g|\leq c+d$ for $i\in [i_0]$
        from the facts that all terminals in $T$ has degree 1 and $|F|\leq c$.
        Thus, $A_i^{\textsf g}\cup \{\sink_i(\bar e)\}$ satisfies the truncation condition of $\mathcal M_i$ for any $i\in[0,i_0]$.
        In the following, we fix an index $i\in [0,i_0]$ and demonstrate that $\{\sink_i(\bar e)\}$ extends $A_i^\textsf{g}$ in the hyperedge gammoid of $(G,T)$.
        Recall that any minimum multiway cut of $\mathcal T$ contains $\bar e$, and $F$ is a minimum multiway cut of $\mathcal T$ in which $|F|\leq c$.

        We show that there are $|F|+1$ pairwise edge-disjoint paths from $T\setminus V_i$ to $F$ in $G$ if we allow using $\bar e\in F$ twice.
        This implies that $\{\sink_i(\bar e)\}\cup A_i^{\textsf g}$ and $A_i^{\textsf g}$ are independent in $\mathcal M_i$ by Observation~\ref{obs:hypergammoid} since we set $A_i^{\textsf{g}}=\{e_i\mid e\in F\cup\partial(T\cap V_i)\}$.
        Furthermore, it implies that $\{\sink_i(\bar e)\}$ extends $A_i^{\textsf g}$ in $\mathcal M_i$.
        We suppose that there are no such $|F|+1$ edge-disjoint paths from $T\setminus V_i$ to $F$.
        By the Menger's theorem for hypergraphs~\cite{zykov1974hypergraphs}, there exists another hyperedge set $F'$ separating $T\setminus V_i$ and $F$ excluding $e$ with $|F'|\leq |F|$. 
        Additionally, any path between two terminals in different components in $\mathcal T$ contains a hyperedge in $F$, consequently, it also contains a hyperedge in $F'$.
        Thus, $F'$ is a minimum multiway cut of $\mathcal T$ excluding $\bar e$.
        This contradicts that $\bar e$ is essential. 
        In conclusion, $\{\sink_i(\bar e)\}$ extends $A_i^{\textsf{g}}$ in $\mathcal M_i$ for any $i\in [0,i_0]$.

        \subparagraph*{Assertion(iii).}

        We establish that any set $J(e)$ for a hyperedge $e \neq \bar e$ cannot extend $A=A^{\textsf u} \cup (\cup_{i \in [0, i_0]} A_i^\textsf g)$ in $\mathcal M$. 
        There are two cases: $e$ is in $E_{\smal}\cup F$ or $e$ is in $E(V_i)$ for some $j\in[0,i_0]$.
        Here, $E_\textsf{small}$ is the union of $E(V_{i_0+1}), E(V_{i_0+2}),\ldots, E(V_s)$.
        In the first case, $e^{\textsf u}\in A^{\textsf u}$ by construction, and thus, $\{e^{\textsf u}\}$ cannot extend $A_{\textsf u}$ in any matroid. Therefore, $J(e)$ cannot extend $A$ in $\mathcal M$.
        
        Consider the latter case that 
        $e$ is in $E(V_i)$ for some index $i \in [0, i_0]$.         
        In this scenario, $\{\sink_i(e)\}$ extends $A_i^\textsf g$ in $\mathcal M_i$ only if 
        there is a path from $T\setminus V_i$ to $e$ excluding $F$  by Observation~\ref{obs:hypergammoid}.
        However, there is no such path in $G\setminus F$ since $V_i\supset e$ is not connected to $T \setminus V_i$ in $G\setminus F$. Thus, $\{\sink_i(e)\}$ cannot extend $A_i^\textsf{g}$ in $\mathcal M_i$, consequently, $J(e)$ cannot extend  $A$ n $\mathcal M$.   
        \end{proof}
        Lemma~\ref{lem:base_exist} holds by Lemma~\ref{lem:representative_set} and Lemma~\ref{lem:existence_corr}.
\begin{restatable}{lemma}{lemBaseExist}\label{lem:base_exist}
    If $(G,T,c)$ is $d$-unbreakable and $\alpha$-dense  with $\alpha\geq 35r\log d$, $c\leq d\leq k$, and $m>kd^{\alpha-1}$, then there is a non-essential hyperedge.
\end{restatable}
\begin{proof}
        It is sufficient to demonstrate that  $|\mathcal J^*|$ is at most $kd^{\alpha-1}$ by Lemma~\ref{lem:existence_corr}. 
        According to Lemma~\ref{lem:representative_set}, the size of $\mathcal J^*$ is bounded by the product of the ranks of the uniform matroid and $\mathcal M_i$'s. Therefore, we have the following from $c\leq d\leq k$, $\kappa=(d/2)^{\alpha-i_0-1}$, and  $\alpha\geq 3i_0+6$.
        \begin{align*}
            |\mathcal J^*|\leq (k+c+1)(d+c+1)^{i_0}\cdot (\kappa+c) 
            \leq 3^{i_0+1}\cdot kd^{i_0+1}\cdot \left(\frac {d} 2\right)^{\alpha-i_0-2}  \leq kd^{\alpha-1}. 
        \end{align*}
        In conclusion, if $m>kd^{\alpha-1}$, a hyperedge is not in $\mathcal J^*$ which is non-essential by Lemma~\ref{lem:existence_corr}.
\end{proof}

\subsection{Non-Essential Hyperedge in a General Instance}\label{sec:exi_unbr_sparse}
In this section, we show that a minimal multicut-mimicking network of $(G,T,c)$ has at most $|T|c^{O(r\log c)}$ hyperedges. 
For this achievement, we start by assuming that $(G,T,c)$ has more than $|T|c^{\Omega(r\log c)}$ hyperedges, and find a subinstance that has a non-essential hyperedge. 
Note that the obtained hyperedge is also non-essential in $(G, T,c)$ by Lemma~\ref{lem:essential_in_subinstance}, and thus, $(G, T,c)$ is not minimal by Lemma~\ref{lem:contracting}. 

First, we find a $5c$-unbreakable subinstance $(G',T',c)$ of $|T'|c^{\Omega(r\log c)}$ hyperedges.
Then we recursively find a subinstance of a $(5c)$-unbreakable instance until it satisfies the conditions in Lemma~\ref{lem:base_exist}.
It has a non-essential hyperedge which is non-essential in the original instance.

\subparagraph*{Construction of unbreakable subinstance.}
We suppose that $(G,T,c)$ is an instance with $m>5|T|\cdot\beta(c)$, where $\beta(c)=(5c)^{35(r+2)\log(5c)-1}$.
Let $r$ and $m$ be the rank and the number of hyperedges, respectively, of $G$.
Our goal is computing a $5c$-unbreakable subinstance $(G',T',c)$ of $(G,T,c)$ such that it has more than $|T'|\cdot \beta(c)$ hyperedges and $|T'|\geq 5c$.
For this, we construct a vertex partition $(V_1,\ldots, V_{s})$ in a greedy manner so that the subinstances corresponding to the components
are $5c$-unbreakable.

We recursively construct a vertex partition starting from the initial partition $(V(G))$ 
until every component of the partition corresponds to a $5c$-unbreakable subinstance. 
Assume that we have a partition $(V_1, \dots, V_{s'})$ at some point such that the subinstance $(\hat G[V_i], T_i, c_i)$ with respect to some component $V_i$ is not $5c$-unbreakable.
Recall that $c_i=\min\{c,|T_i|\}$ and $T_i$ is the union of $T\cap V_i$ and anchored terminals.
By the definition of the unbreakable property,
there exists a vertex subset $X$ in $\hat G[V_i]$ with $|\partial_{\hat G[V_i]} X|\leq c_i$ such that $|X\cap T_i|\geq 5c$ and $|T_i\setminus X|\geq 5c$. 
Then we decompose $V_i$ into $V_i\cap X$ and $V_i\setminus X$ in the current partition. 
Notice that every subinstance in the partition contains at least $5c$ terminals in $T_i$ at any point by construction.
Thus, the subinstances preserve $c_i=c$ due to $c_i=\min\{c,|T_i|\}$.

For the finally obtained vertex partition $(V_1,\dots,V_s)$, we demonstrate that $\sum_{i\in [s]}|T_i|$ is at most $5|T|$.
The number of decomposing iterations called is equal to $s-1$, where $s$ denotes the size of the returned partition. 
Thus the following inequality holds which implies that $s$ is at most $|T|/c$, and $\sum_{i\in [s]}|T_i|\leq 5|T|$:
    \begin{align*}
        5cs\leq \sum_{i\in [s]}|T_i|\leq |T|+4cs.
    \end{align*}
    The first inequality is obtained from $|T_i|\geq 5c$ for any $i\in [s]$. 
    Additionally, the last one holds since each separating iteration generates at most $4c$ (anchored) terminals.

For the obtained vertex partition $(V_1,\dots, V_s)$, at least one subinstance $(\hat G[V_i],T_i,c)$ has more than $|T_i|\beta(c)$ hyperedges.
This is because if every instances has at most $|T_i|\beta(c)$ hyperedges, then $G$ has at most $\sum_{i\in[s]}|T_i|\beta(c)\leq 5|T|\beta(c)$ hyperedges, which contradicts for the assumption that $m>5|T|\beta(c)$, where $m=|E(G)|$.

In conclusion, if  $(G,T,c)$ has more than $5|T|\beta(c)$ hyperedges, then it has a subinstance $(\hat G[V_i],T_i,c)$ which is $5c$-unbreakable and has more than $|T_i|\beta(c)$ hyperedges, where $\beta(c)=(5c)^{35(r+2)\log(5c)-1}$.
In the rest of this section, we focus on a $(5c)$-unbreakable instance $(G,T,c)$ with more than $|T|c^{\Omega(r\log c)}$ hyperedges.
However, it increases the hyperedge size only if the hyperedge has less than two terminals.
Additionally, once increased hyperedge has two (anchored) terminals.
Thus, the rank is increased by at most one even if we obtain a subinstance recursively.

\medskip

In the following, we recursively find a subinstance of a $(5c)$-unbreakable instance until it satisfies the conditions in Lemma~\ref{lem:base_exist}.
It has a non-essential hyperedge, furthermore, it is also non-essential in the original instance.
Note that constructing a subinstance might increase {the size of a hyperedge} by inserting two anchored terminals for a restricted hyperedge.
In the remainder, we fix $r$ as the rank of the original instance to avoid confusion.  

\subparagraph*{Construction of non-minimal instance.} 
We suppose that $(G,T,c)$ is $d$-unbreakable with $d=\min\{5c,|T|\}$. 
Then we show that if $G$ has more than $|T|d^{\alpha(c)-1}$ hyperedges, $(G,T,c)$ has a non-essential hyperedge by inductively along $m$, where $\alpha(c)=35(r+2)\log (5c)$.
Recall that $r$ is a {fixed} constant so that the rank of $(G, T,c)$ is at most $r+1$, and $\alpha$ is the constant derived from $c$ only.
For simplicity, we use $\alpha$ to denote {$\alpha(c)$} when the context is clear.
If $(G,T,c)$ is  $\alpha$-dense, then it has a non-essential hyperedge by Lemma~\ref{lem:base_exist}.

When $(G,T,c)$ is not $\alpha$-dense, there is a witness vertex set $X\subseteq V$ with 
 {$0<|E(X)|\leq|E(V\setminus X)|$}, $|\partial X|\leq c$, and  {$|E(X)|>(\capacity_T(X))^\alpha$}.\footnote{$(G,T,c)$ is $\alpha$-dense if 
    $ |E(Y)|\leq (\capacity_T(Y))^{\alpha} $ for any $Y\subseteq V$ with $0<|E(Y)|\leq |E(V\setminus Y)|$ and $|\partial Y|\leq c$.} 
If $|X\cap T|>|T\setminus X|$, we \emph{replace} $X$ with $V\setminus X$. 
Note that the following inequalities still hold: 
$|\partial X|\leq c$, {$|E(X)|>(\capacity_T(X))^\alpha$}, and $|E(V\setminus X)|>0$. 
The first one holds since $\partial X=\partial (V\setminus X)$, and the second one holds since $|T\cap X|$ and $\capacity_T(X)$ are decreased by the replacement while $|E(X)|$ is increased.
The last holds since we chose $X$ so that $E(X), E(V\setminus X)\neq \emptyset$.
Additionally, the size of $T\cap X$ is at most $d$ since our instance is $d$-unbreakable.
We move to the subinstance $(\hat G[X],T_X, c_X)$ with respect to $X$, where $c_X= \min\{c,|T_X|\}$ and $T_X$ is the union of terminals $T\cap X$ and the anchored terminals. 
Recall that the size of $T_X$ is at most $\capacity_T(X;G)=|T\cap X|+2|\partial X|\leq d+2c$.
Lemma~\ref{lem:condition_hold} ensures the safeness of this inductive proof.

\begin{restatable}{lemma}{LemConditionHold}\label{lem:condition_hold}
    $(\hat G[X],T_X,c_X)$ is $d_X$-unbreakable with $d_X=\min\{5c_X,|T_X|\}$.
    Additionally, $\hat G[X]$ has more than $|T_X|d_X^{\alpha'-1}$ hyperedges but less than $|E(G)|$, where $\alpha'=\alpha(c_X)$.
\end{restatable}
\begin{proof}
We first show that $(\hat G[X],T_X,c_X)$ is $d_X$-unbreakable.
For this, we show that the size of $T_X$ is at most $2d_X$, this implies that $(\hat G[X],T_X,c_X)$ is  $d_X$-unbreakable since the instance is $|T_X|/2$-unbreakable clearly.\footnote{For any $Y\subset V(G)$, $|T\cap Y|$ or $|T\setminus Y|$ is at most $|T|/2$. }
Recall that the size of $T_X$ is at most $\capacity_T(X;G)=|T\cap X| + 2|\partial X|$.
Therefore, the size of $T_X$ is at most $d+2c$ since we chose $X$ so that the size of $\partial X$ and $T\cap X$ are at most $c$ and $d$, respectively. 
Note that we have $d+2c\leq 7c$ from $d=\min \{5c,|T|\}$.
Furthermore, this implies that $|T_X|<2d_X$  from $d_X=\min\{5c_X,|T_X|\}$ with $c_X=\min\{c,|T_X|\}$, and thus, $(\hat G[X],T_X,c_X)$ is  $d_X$-unbreakable.


Recall that we chose $X$ so that $E(V\setminus X)$ is not empty and {$|E(X)|>(\capacity_T(X;G))^{\alpha(c)}$}.
It is trivial that $\hat G[X]$ has strictly less than $m$ hyperedges.
Furthermore, $\hat G[X]$ has  more than $|T_X|d_X^{\alpha'-1}$ hyperedges.
Precisely, we have 
    \begin{align*}
        |E(\hat G[X])|>|T_X|^{\alpha}\geq |T_X|d_X^{\alpha'-1}.
    \end{align*}
    Recall that $d_X\leq |T_X|\leq \capacity_T(X;G)$ and $\alpha'=\alpha(c_X)\leq \alpha(c)$ by $c_X\leq c$.
\end{proof}

We recursively obtain a subinstance until it becomes $\alpha$-dense. 
Note that $k, d, m$, and $\alpha$ change during the recursion while the rank is always at most $r+1$, where $k$ and $m$ denote the number of terminals and hyperedges, respectively, in the current instance.
However, Lemma~\ref{lem:condition_hold} guarantees that $m>kd^{\alpha-1}$ holds at each step.
Moreover, $m$ is strictly decreased.
Thus, we always reach a $d$-unbreakable and $\alpha$-dense instance satisfying the conditions of Lemma~\ref{lem:base_exist}, and it has a non-essential hyperedge.
It is easy to show that the hyperedge is also non-essential in the original instance by applying Lemma~\ref{lem:essential_in_subinstance} recursively.

In conclusion, a $d$-unbreakable instance $(G, T,c)$ with $m>|T|d^{\alpha(c)-1}$ and $d=\min\{5c,|T|\}$ has a non-essential hyperedge.
This section proves Theorem~\ref{thm:exists}.
\thmExists*

\section{Conclusion and Further Works}\label{sec:conclu_ap}
We demonstrate that if a hypergraph instance $(G,T,c)$ has more than $|T|c^{O(r\log c)}$ hyperedges, then we can always get a smaller-sized multicut-mimicking network by contracting a hyperedge.
Furthermore, we propose an efficient algorithm to find it.
In conclusion, we introduce an efficient algorithm that constructs a multicut-mimicking network with $|T|c^{O(r\log c)}$ hyperedges. As a natural next step, further exploration could focus on reducing the time complexity or the size of the returned multicut-mimicking network. 
For instance, is there a multicut-mimicking network of size $|T|c^{O(\log (rc))}$?
Additionally, investigating further variations of vertex sparsifiers, such as considering trade-offs between the network size, running time, and solution quality (approximating factor), would be valuable in future research. 

\subparagraph{Approximated sparsification.}
Graph sparsification has an approximated version that reduces the size of a graph (or hypergraph) while maintaining the objective function value up to a multiplicative factor $q$, called \emph{quality}. 
For example, our multicut-mimicking network research was focused on $q=1$.
Additionally, there are previous researches focused on achieving quality bounds $q\in O(\log |T|/\log\log|T|)$ and lower bound $q\in \Omega(\sqrt{\log |T|}/\log\log|T|)$ for the $2$-way cut problem in graphs~\cite{charikar2010vertex,leighton2010extensions,makarychev2010metric,moitra2009approximation}.
Investigating the trade-offs between the solution quality, network size, and running time for multiway cuts would be valuable.



\bibliography{paper}

\begin{thebibliography}{10}

\bibitem{alpert1995recent}
Charles~J. Alpert and Andrew~B. Kahng.
\newblock Recent directions in netlist partitioning: A survey.
\newblock {\em Integration}, 19(1-2):1--81, 1995.

\bibitem{bansal2019new}
Nikhil Bansal, Ola Svensson, and Luca Trevisan.
\newblock New notions and constructions of sparsification for graphs and hypergraphs.
\newblock In {\em Proceedings of the 60th Annual Symposium on Foundations of Computer Science (FOCS 2019)}, pages 910--928, 2019.

\bibitem{soda2021vertexspar}
Parinya Chalermsook, Syamantak Das, Yunbum Kook, Bundit Laekhanukit, Yang~P. Liu, Richard Peng, Mark Sellke, and Daniel Vaz.
\newblock Vertex sparsification for edge connectivity.
\newblock In {\em Proceedings of the 32nd ACM-SIAM Symposium on Discrete Algorithms (SODA 2021)}, pages 1206--1225, 2021.

\bibitem{charikar2010vertex}
Moses Charikar, Tom Leighton, Shi Li, and Ankur Moitra.
\newblock Vertex sparsifiers and abstract rounding algorithms.
\newblock In {\em Proceedings of the 51st Annual Symposium on Foundations of Computer Science (FOCS 2010)}, pages 265--274, 2010.

\bibitem{chechik2018near}
Shiri Chechik and Christian Wulff-Nilsen.
\newblock Near-optimal light spanners.
\newblock {\em ACM Transactions on Algorithms (TALG)}, 14(3):1--15, 2018.

\bibitem{chekuri2018minimum}
Chandra Chekuri and Chao Xu.
\newblock Minimum cuts and sparsification in hypergraphs.
\newblock {\em SIAM Journal on Computing}, 47(6):2118--2156, 2018.

\bibitem{chen20241+}
Yu~Chen and Zihan Tan.
\newblock On $(1+\varepsilon)$-approximate flow sparsifiers.
\newblock In {\em Proceedings of the 35th Annual ACM-SIAM Symposium on Discrete Algorithms (SODA 2024)}, pages 1568--1605, 2024.

\bibitem{cygan2015parameterized}
Marek Cygan, Fedor~V. Fomin, {\L}ukasz Kowalik, Daniel Lokshtanov, D{\'a}niel Marx, Marcin Pilipczuk, Micha{\l} Pilipczuk, and Saket Saurabh.
\newblock {\em Parameterized algorithms}, volume~4.
\newblock Springer, 2015.

\bibitem{dahlhaus1994complexity}
Elias Dahlhaus, David~S. Johnson, Christos~H. Papadimitriou, Paul~D. Seymour, and Mihalis Yannakakis.
\newblock The complexity of multiterminal cuts.
\newblock {\em SIAM Journal on Computing}, 23(4):864--894, 1994.

\bibitem{filtser2016greedy}
Arnold Filtser and Shay Solomon.
\newblock The greedy spanner is existentially optimal.
\newblock In {\em Proceedings of the 35th ACM Symposium on Principles of Distributed Computing (PODC 2016)}, pages 9--17, 2016.

\bibitem{jiang2022hypergraphvertex}
Han Jiang, Shang-En Huang, Thatchaphol Saranurak, and Tian Zhang.
\newblock Vertex sparsifiers for hyperedge connectivity.
\newblock In {\em Proceedings of the 30th Annual European Symposium on Algorithms (ESA 2022)}, pages 70:1--70:13, 2022.

\bibitem{jin2022fully}
Wenyu Jin and Xiaorui Sun.
\newblock Fully dynamic $s$-$t$ edge connectivity in subpolynomial time.
\newblock In {\em Proceedings of the 62nd Annual Symposium on Foundations of Computer Science (FOCS 2022)}, pages 861--872. IEEE, 2022.

\bibitem{kappes2016higher}
J{\"o}rg~Hendrik Kappes, Markus Speth, Gerhard Reinelt, and Christoph Schn{\"o}rr.
\newblock Higher-order segmentation via multicuts.
\newblock {\em Computer Vision and Image Understanding}, 143:104--119, 2016.

\bibitem{kapralov2022spectral}
Michael Kapralov, Robert Krauthgamer, Jakab Tardos, and Yuichi Yoshida.
\newblock Spectral hypergraph sparsifiers of nearly linear size.
\newblock In {\em Proceedings of the 62nd Annual Symposium on Foundations of Computer Science (FOCS 2022)}, pages 1159--1170, 2022.

\bibitem{kratsch2012representative}
Stefan Kratsch and Magnus Wahlstr{\"o}m.
\newblock Representative sets and irrelevant vertices: New tools for kernelization.
\newblock {\em Journal of the ACM}, 67(3):1--50, 2020.

\bibitem{krauthgamer2023exact}
Robert Krauthgamer and Ron Mosenzon.
\newblock Exact flow sparsification requires unbounded size.
\newblock In {\em Proceedings of the 34th Annual ACM-SIAM Symposium on Discrete Algorithms (SODA 2023)}, pages 2354--2367, 2023.

\bibitem{leighton2010extensions}
F~Thomson Leighton and Ankur Moitra.
\newblock Extensions and limits to vertex sparsification.
\newblock In {\em Proceedings of the 42th ACM Symposium on Theory of computing (STOC 2010)}, pages 47--56, 2010.

\bibitem{liu2023vertex_poly}
Yang~P Liu.
\newblock Vertex sparsification for edge connectivity in polynomial time.
\newblock In {\em Proceedings of the 14th Innovations in Theoretical Computer Science Conference (ITCS 2023)}, pages 83:1--83:15, 2023.

\bibitem{lokshtanov2018deterministic}
Daniel Lokshtanov, Pranabendu Misra, Fahad Panolan, and Saket Saurabh.
\newblock Deterministic truncation of linear matroids.
\newblock {\em ACM Transactions on Algorithms (TALG)}, 14(2):1--20, 2018.

\bibitem{long2022near}
Yaowei Long and Thatchaphol Saranurak.
\newblock Near-optimal deterministic vertex-failure connectivity oracles.
\newblock In {\em Proceedings of the 63rd Annual Symposium on Foundations of Computer Science (FOCS 2022)}, pages 1002--1010, 2022.

\bibitem{louis2014approximation}
Anand Louis and Yury Makarychev.
\newblock Approximation algorithms for hypergraph small set expansion and small set vertex expansion.
\newblock {\em Approximation, Randomization, and Combinatorial Optimization. Algorithms and Techniques}, page 339, 2014.

\bibitem{makarychev2010metric}
Konstantin Makarychev and Yury Makarychev.
\newblock Metric extension operators, vertex sparsifiers and lipschitz extendability.
\newblock In {\em Proceedings of the 51st Annual Symposium on Foundations of Computer Science (FOCS 2010)}, pages 255--264, 2010.

\bibitem{marx2006parameterized}
D{\'a}niel Marx.
\newblock Parameterized graph separation problems.
\newblock {\em Theoretical Computer Science}, 351(3):394--406, 2006.

\bibitem{marx2009parameterized}
D{\'a}niel Marx.
\newblock A parameterized view on matroid optimization problems.
\newblock {\em Theoretical Computer Science}, 410(44):4471--4479, 2009.

\bibitem{moitra2009approximation}
Ankur Moitra.
\newblock Approximation algorithms for multicommodity-type problems with guarantees independent of the graph size.
\newblock In {\em Proceedings of the 50th Annual Symposium on Foundations of Computer Science (FOCS 2009)}, pages 3--12, 2009.

\bibitem{oxley2022matroid}
James Oxley.
\newblock Matroid theory.
\newblock In {\em Handbook of the Tutte Polynomial and Related Topics}, pages 44--85. Chapman and Hall/CRC, 2022.

\bibitem{ozdal2004hypergraph}
Muhammet~Mustafa Ozdal and Cevdet Aykanat.
\newblock Hypergraph models and algorithms for data-pattern-based clustering.
\newblock {\em Data Mining and Knowledge Discovery}, 9:29--57, 2004.

\bibitem{wahlstrom2022quasipolynomial_multicut}
Magnus Wahlstr{\"o}m.
\newblock Quasipolynomial multicut-mimicking networks and kernels for multiway cut problems.
\newblock {\em ACM Transactions on Algorithms (TALG)}, 18(2):1--19, 2022.

\bibitem{zhang2010hypergraph}
Zi-Ke Zhang and Chuang Liu.
\newblock A hypergraph model of social tagging networks.
\newblock {\em Journal of Statistical Mechanics: Theory and Experiment}, 2010(10):P10005, 2010.

\bibitem{zykov1974hypergraphs}
Alexander~Aleksandrovich Zykov.
\newblock {\em Hypergraphs}, volume~29.
\newblock IOP Publishing, 1974.

\end{thebibliography}
\end{document}